\documentclass[conference]{IEEEtran}
\IEEEoverridecommandlockouts
\usepackage{cite}
\usepackage{url}      
\usepackage{amsmath,amssymb,amsfonts}

\usepackage{graphicx}
\usepackage{textcomp}
\usepackage{xcolor}
\def\BibTeX{{\rm B\kern-.05em{\sc i\kern-.025em b}\kern-.08em
    T\kern-.1667em\lower.7ex\hbox{E}\kern-.125emX}}

\usepackage{amsthm}
\usepackage[normalem]{ulem} 
\usepackage[table]{xcolor} 

\usepackage{cases}
\makeatletter
\def\th@newremark{\th@remark\thm@headfont{\bfseries}}
\makeatletter
\theoremstyle{newremark}
\newtheorem{definition}{Definition}

\usepackage{tikz}

\usepackage[ruled]{algorithm}
\PassOptionsToPackage{noend}{algpseudocode}
\usepackage{algpseudocode}
\usepackage{amsmath}
\usepackage{bm}
\usepackage{color}
\usepackage{etex}
\usepackage{qtree}
\usepackage{graphicx}
\usepackage{multirow}
\usepackage{multicol}
\usepackage{subfigure}
\usepackage{url}
\usepackage{thmtools}
\usepackage{enumitem}

\usepackage{tabularx}
\usepackage{balance}    
\usepackage{booktabs}   
\usepackage{graphics}   
\usepackage{url}        
\usepackage{pifont}     

\usepackage{xcolor}      
\usepackage[hidelinks]{hyperref}  
\usepackage{xspace}
\usepackage{balance}
\newcommand{\ie}[0]{\textit{i.e.},\ }   
\newcommand{\eg}[0]{\textit{e.g.},\ }   
\newcommand{\etc}[0]{\textit{etc.}\ }   

\usepackage{pifont}
\usepackage{ctable}

\renewcommand{\algorithmiccomment}[1]{\bgroup\hfill//~#1\egroup}

\newcommand{\eat}[1]{}

\definecolor{darkgreen}{rgb}{0.0, 0.5, 0.0}

\definecolor{ballblue}{rgb}{0.13, 0.67, 0.8}

\newcommand{\app}{\textsc{Krone}}

\newcommand{\deeplog}{\textit{Deeplog}}
\newcommand{\loganomaly}{\textit{LogAnomaly}}
\newcommand{\logrobust}{\textit{LogRobust}}
\newcommand{\logbert}{\textit{LogBert}}
\newcommand{\logprompt}{\textit{LogPrompt}}

\newcommand{\appseq}{\textit{\app{} Seq}}
\newcommand{\appseqs}{\textit{\app{} Seqs}}

\newcommand{\entityseq}{\textit{\app{} E-seq}}
\newcommand{\entityseqs}{\textit{\app{} E-seqs}}
\newcommand{\actionseq}{\textit{\app{} A-seq}}
\newcommand{\actionseqs}{\textit{\app{} A-seqs}}
\newcommand{\statusseq}{\textit{\app{} S-seq}}
\newcommand{\statusseqs}{\textit{\app{} S-seqs}}

\newcommand{\datamodel}{\textit{\app{} Log Abstraction Model}}

\newcommand{\atomic}{\textit{Modularity}}
\newcommand{\horizontal}{\textit{Hierarchical Locality}}
\newcommand{\vertical}{\textit{Nested Semantics}}

\newcommand{\localformat}{Local-Context}
\newcommand{\nestedformat}{Nested-Aware}

\newcommand{\partitionalgo}{\textsc{GenerateKroneSeq}}
\newcommand{\algo}{\textsc{TopDownSeqDecompose}}

\newcommand{\commentsymbol}{/*}
\algrenewcommand\algorithmiccomment[1]{\hfill \commentsymbol{} #1}

\definecolor{myblue}{RGB}{135,206,250} 
\definecolor{lighterorange}{RGB}{255,225,180}  
 \definecolor{lightorange}{RGB}{255,200,130}
\definecolor{lightgray}{RGB}{240,240,240}  

\usepackage{booktabs}

\usepackage{tikz}
\usepackage{xcolor}

\newcommand{\coloredcircled}[2]{%
    \tikz[baseline=(char.base)]{
        \node[shape=circle,draw,fill=#1,inner sep=0.6pt] (char) {\textbf{#2}};
    }
}

\definecolor{entityblue}{RGB}{15, 158, 213}
\definecolor{actiongreen}{RGB}{78, 167, 46}
\definecolor{statusorange}{RGB}{233, 113, 50}
\begin{document}

\title{\app{}: Scalable LLM-Augmented Log Anomaly Detection via Hierarchical Abstraction\thanks{Accepted at ICDE 2026.}
}

\author{
\IEEEauthorblockN{
Lei Ma\IEEEauthorrefmark{1},
Jinyang Liu\IEEEauthorrefmark{2},
Tieying Zhang\IEEEauthorrefmark{2},
Peter M. VanNostrand\IEEEauthorrefmark{1},
Dennis M. Hofmann\IEEEauthorrefmark{1},\\
Lei Cao\IEEEauthorrefmark{3},
Elke A. Rundensteiner\IEEEauthorrefmark{1},
Jianjun Chen\IEEEauthorrefmark{2}
}

\IEEEauthorblockA{
\IEEEauthorrefmark{1}Worcester Polytechnic Institute, Worcester, USA\\
\IEEEauthorrefmark{2}ByteDance Inc., San Jose, USA\\
\IEEEauthorrefmark{3}University of Arizona, Tucson, USA\\
}
}

\maketitle

\begin{abstract}

Log anomaly detection is crucial for uncovering system failures and security risks. Although logs originate from nested component executions with clear boundaries, this structure is lost when 
stored as flat sequences. 
Hence,  state-of-the-art methods risk missing true dependencies within executions while learning spurious ones across unrelated events. 
We propose \app{}, the first hierarchical anomaly detection framework that automatically derives execution hierarchies from flat logs for modular multi-level anomaly detection.   At its core, the \datamodel{} models log data by extracting the application-specific semantic hierarchical structure. This
hierarchy is then leveraged by \app{} to recursively decompose log sequences into multi-levels of coherent execution chunks, \ie \appseqs{}, transforming sequence-level detection into a set of modular \appseq{}-level detection tasks. For each test \appseq{}, \app{} adopts a hybrid modular detection mechanism that routes between an efficient level-independent \localformat{} detector that rapidly filters normal \appseqs{}, and a \nestedformat{} detector that incorporates cross-level semantic dependencies; augmented with LLM-based anomaly detection and explanation.  
\app{} optimizes the modular detection tasks along the hierarchy with cached result reuse and early-exit optimization strategies. 
Experiments on three public benchmarks and one industrial dataset from ByteDance Cloud demonstrate the comprehensive improvement of \app{}, on accuracy (42.49\% $\rightarrow$ 87.98\%), F-1 of same detector with or without hierarchy), data-efficiency (data space 117.3$\times\downarrow$), resource-efficieny (43.7$\times\downarrow$) and interpretability.
\app{} improves F1-score by 10.07\% (82.76\% $\rightarrow$ 92.83\%) over prior methods, while reducing LLM usage to 1.1\%–3.3\% of the test data size. The code is available at \url{https://github.com/LeiMa0324/Krone_official}. We also provide an interactive demo at: \url{https://leima0324.github.io/KRONE_Demo_official/}.

\end{abstract}


\section{Introduction}
\label{sec:intro}

The explosive growth of complex engineered systems 
has made detecting anomalies from massive log data critical for preventing 
system failures and cyber intrusions. By learning 
patterns of normal log sequences, most deep-learning methods~\cite{deeplog, OC4SEQ, RT-log,logbert,meng2019loganomaly, logrobust, cat, pluto, le2021log, logsy} use model architectures such as LSTM, RNN, or Transformer, to detect abnormal log sequences based on their deviation from the learned normal patterns. These methods, consuming flat log sequences, tend to overlook the semantics-rich \textit{hierarchical}
execution patterns hidden
in these sequences. 
As the motivating example below illustrates, recognizing and leveraging  these patterns 
in log sequences offers unique opportunities for log analysis. 
%
%
For ease of exposition,
we
illustrate 
the hierarchical log abstraction 
 core for empowering anomaly detection 
 on simple log data, our method is rigorously evaluated 
 on complex real-world commercial logs (Section~\ref{sec:exp}).

\begin{figure}[t]
    \centering
    \includegraphics[width=0.99\linewidth]{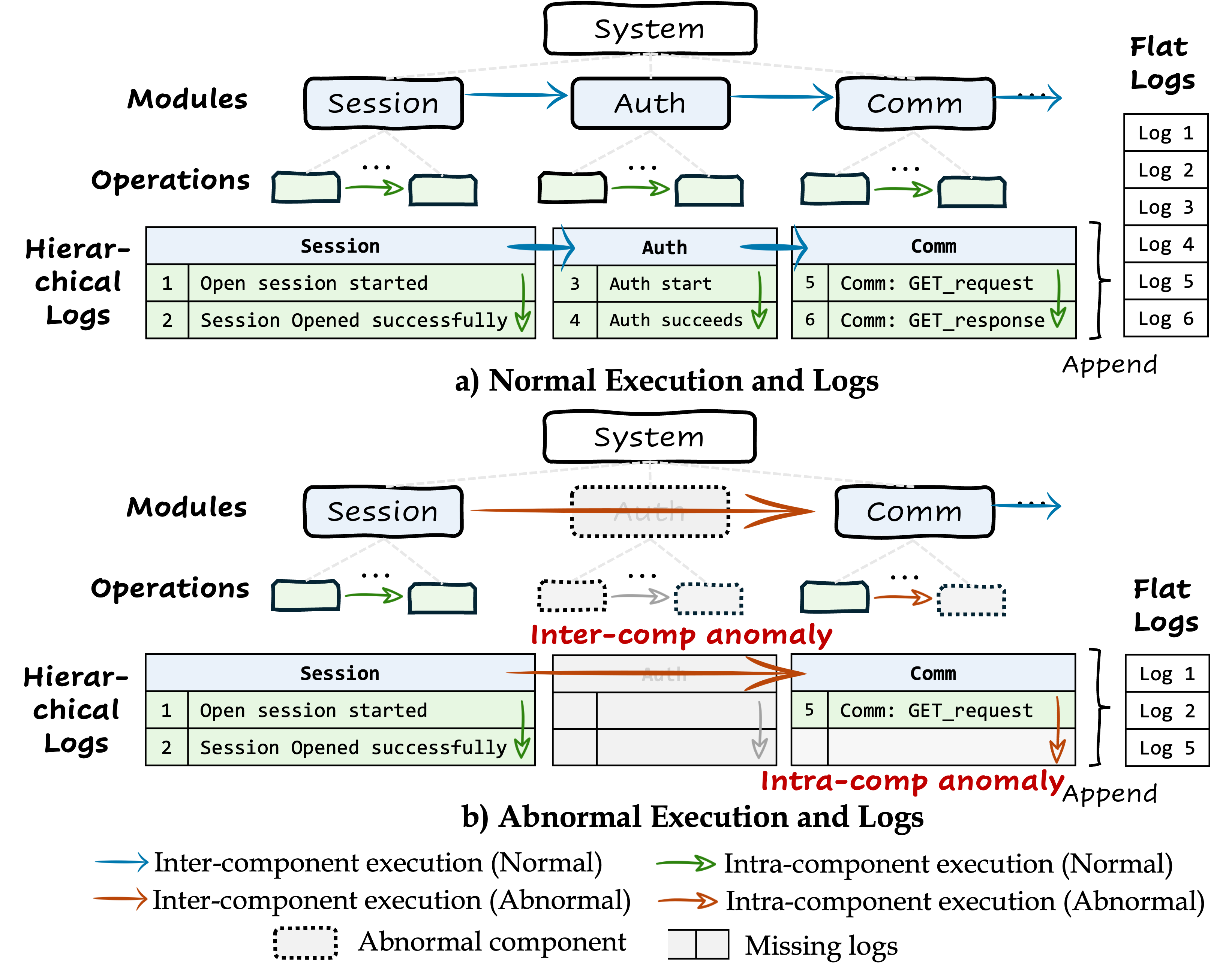}
    \vspace{-20pt}
    \caption{Hierarchical System, Execution,  and Log Anomalies. Low-level anomaly (intra-component): missing response for $\texttt{GET}\_request$ in $Comm$ component. High-level anomaly (inter-component):  abnormal component transition $[Session, Comm]$ without $Auth$.}
    \label{fig:intro}
    \vspace{-20pt}
\end{figure}

\noindent\textbf{Motivating Example.} Figure~\ref{fig:intro} (top) illustrates a 
hierarchical software system whose components 
produce 
logs representing 
 nested execution units.
%
The login system consists of three components  $Session$, $Auth$, and $Comm$, each with its own operations (\eg classes, functions). A execution follows the inter-component order $[Session, Auth, Comm]$ (blue arrows), with each component generating its own intra-component logs (green arrows): Logs 1–2 from $Session$, 3–4 from $Auth$, and 5–6 from $Comm$, with clear execution boundaries. 

We posit that such execution boundaries reveal a natural hierarchical organization of system executions. By leveraging these boundaries, we can \textit{decompose complex executions into reusable coherent patterns}—where each intra-component segment represents an atomic execution unit and inter-component sequences encode their permutations. Under this 
nested view, log sequences are no longer treated as opaque lengthy traces but instead as structured compositions of atomic patterns. 

\noindent\textbf{Benefits for Anomaly Detection.}
This hierarchical organization of log data yields two key advantages for accurate, scalable and intepretable anomaly detection.

\textit{Modular Reusable Learning and Detection. }
Instead of relying solely on 
sequence models to capture long-range dependencies~\cite{deeplog,logbert,meng2019loganomaly,logrobust}, we observe that each component typically exhibits only a small set of recurring atomic behavioral patterns, whereas the vast diversity of log sequences arises largely from their permutations. By learning and detecting these shorter reusable atomic patterns and their permutations, we demonstrate that detection can become simpler, more accurate, and data-efficient
(See results in Section \ref{sec:exp:hier_viz}).

\textit{Hierarchical Interpretation of Anomalies. } 
This organization of log data naturally yields a multi-level view of anomalies. As illustrated in Figure~\ref{fig:intro}  (bottom), erroneous deviations may occur within a component (intra-component anomalies) or  across components 
with deviated permutation (inter-component anomalies). Modeling anomalies at their 
level of abstraction
improves localization and interpretability 
(See experimental results in Section \ref{sec:exp:hier_viz}, Figure \ref{fig:hier_anomaly}).

Although a hierarchy offers advantages, real-world logs are typically 
stored and modeled as flat sequences~\cite{logsed}, obscuring their latent hierarchical 
structure. 
We 
thus aim  to enable  accurate, efficient, and interpretable anomaly detection by uncovering  the hidden hierarchical structure in logs.

\noindent\textbf{Challenges.}
Realizing this vision of \textit{modular 
hierarchical anomaly detection} faces several challenges.

\textit{Challenge 1: Effective  Hierarchy Abstraction 
of Black-Box Systems. } 
Software systems typically function as black boxes, 
without explicit structural information available that could be leveraged for analysis.
Thus, an effective hierarchical abstraction must be capable of inferring structure
solely from logs, while still accurately capturing execution semantics and  reducing the data space to enable 
modular
pattern learning and detection. 
Achieving this without domain knowledge remains a fundamental challenge.

\textit{Challenge 2: Lightweight Adaptive 
Modeling under
Massive Logs} 
Real-world logs are  extremely large, 
making heavy or complex modeling impractical. A useful hierarchy abstraction should be lightweight enough to construct efficiently, without relying on manual updating of hard-coded rules. Achieving such lightweight modeling for massive log data remains a key challenge. 

\textit{Challenge 3: Detection Efficiency with 
Nested Semantics.} In hierarchical execution, executions  may share identical high-level component permutations, \eg $[Session, Auth, Comm]$, but 
differ due to different low-level intra-component executions. This nested semantics presents a core trade-off: anomaly detection only at
the high-level is fast but coarse, whereas incorporating lower-level details improves accuracy but incurs higher computational costs.
Balancing semantic fidelity and efficiency is essential for precise yet scalable
detection.

\vspace{0.3em}

\noindent\textbf{State-of-the-Art and Their Limitations.} 
Most  log anomaly detection methods~\cite{deeplog,OC4SEQ,RT-log,logbert,meng2019loganomaly,logrobust,cat,pluto,le2021log,logsy} use deep learning models to model flat log sequences, detecting anomalies as deviations from learned normal patterns. However, this strategy introduces a \textit{flattened-dependency dilemma}: either spend a high cost to model long-range dependencies or use efficient sliding window chunking that risks losing them. Although hierarchy has been explored as a remedy, existing approaches~\cite{HitAnomaly,10070784,LayerLog,HLogformer} remain text-centric: they either consider it as multi-scale textual representation~\cite{HitAnomaly,10070784} via hierarchical Transformer~\cite{nawrot2022hierarchicaltransformersefficientlanguage,pappagari2019hierarchicaltransformerslongdocument}, or treat it as shallow text structures such as word→log→sequence~\cite{LayerLog} or JSON-style fields~\cite{HLogformer}. These designs overlook the critical insight that the execution itself has an inherent semantic hierarchy across system components, where anomalies 
actually occur. DeepTraLog~\cite{DeepTraLog} comes closest to modeling execution by decomposing traces into span events, yet relies on hard-coded rules tied to the TrainTicket system - thus limiting its
generalizability. 

Recent works have explored 
large language models (LLM) for log anomaly detection~\cite{Bridging_the_Gap, Early_Exploration,jin2024largelanguagemodelsanomaly,akhtar2025llmbasedeventloganalysis,zhang2025xraglog,logprompt}.
The \textit{flattened-dependency dilemma} 
now manifests as context-window limits~\cite{wang2024limitssurveytechniquesextend,liu2025comprehensivesurveylongcontext}. 
These methods either apply LLMs only on a single log message~\cite{logprompt} due to the context window limitation and LLM-execution costs, which fails to detect sequence-level anomalies, 
or apply them exhaustively to the entire massive sequences~\cite{jin2024largelanguagemodelsanomaly,Early_Exploration,zhang2025xraglog,Bridging_the_Gap}, 
putting the scalability of these detection systems
into question 
~\cite{akhtar2025llmbasedeventloganalysis}.

\noindent\textbf{Proposed Method.}
We propose \app{}, the first hierarchical anomaly detection framework, 
that automatically derives execution hierarchies from flat logs and leverages this structure for anomaly detection. \app{} introduces a \textit{foundational orchestration framework} for log anomaly detection that decomposes, executes, optimizes, and aggregates modular sub-problems, enabling detectors-as-operators to be systematically coordinated for log sequences.

At the core of \app{} is the \datamodel{} (Section~\ref{sec:data_model}), a simple yet
general hierarchical abstraction for modeling log data that enables automatic discovery of semantic structures, 
addressing Challenge 1. Inspired by the semantic structure of logs—where a status reflects the outcome of an action on an entity—\datamodel{} defines a three-level hierarchy: Entity, Action, and Status. To instantiate this hierarchy per dataset (Challenge 2), we formulate the extraction as a Named Entity Recognition (NER) task~\cite{NER_survey} and adopt an LLM-based approach inspired by GraphRAG~\cite{graphRAG} for its adaptability. Leveraging this hierarchy, \app{} propose \algo{} algorithm, which recursively decomposes each log sequence into multi-level \appseqs{}, \ie sequences of entities, actions and statuses, transforming detection task into a set of modular subtasks on the \appseqs{}. As atomic patterns, artifacts of low-level \appseqs{}, such as embeddings, summaries and predictions, can be reused for reoccurrence or hierarchical composition for detection of higher-level \appseqs{}, benefiting a wide range of detectors.

For each test log sequence, \app{} first decomposes it into test \appseqs{} and then performs modular detection on the \appseqs{} bottom-up along the hierarchy. For modular detection, \app{} introduces two complementary \appseq{}-level strategies: \localformat{} for efficient level-independent detection and \nestedformat{} for accurate semantics-aware detection. We unify them into a hybrid approach that dynamically routes between fast pattern matching to filter known normals and selective LLM reasoning for unseen cases, effectively addressing Challenge 3. Results are cached and reused to avoid redundant computation, and an early-exit policy halts detection as soon as a \appseq{} is classified as abnormal. While these optimizations are general to hierarchical execution and independent of the specific detectors, they yield especially significant savings for scalable LLM-based detection.


\noindent\textbf{Contributions.} \app{} provides the following contributions. 

\begin{enumerate}[leftmargin=6mm,topsep=0mm,itemsep=0mm]
    \item  Propose a  hierarchical abstraction for log data, which can be adaptively  extracted for multi-scale log anomaly detection from any new application data set.

    \item Develop a streamlined (optionally LLM-assisted) pipeline that instantiates the hierarchy dynamically and that way decomposes logs into modular \appseqs{}, achieving up to 117.3$\times$ data-space reduction (Figure~\ref{fig:cadinality}).

    \item Design hybrid \appseq{}-level modular detection with adaptive routing between a lightweight \localformat{} detector of pattern matching and a \nestedformat{} detector using LLM reasoning, to balance efficiency and accuracy. 

    \item Develop resource-optimized \app{} 
    detection framework that
    executes the modular detection tasks along the hierarchy.  Experiments show that even with a simple detector, hierarchical execution brings significant F-1 improvement (42.49\% $\rightarrow$ 87.98\%) over non-hierarchical detection (Table~\ref{tab:comapre_PM}),  with resource savings up to 43.7$\times$(Figure~\ref{fig:tokens}).  

    \item Evaluate \app{} on three public benchmarks and one Bytedance industrial dataset—revealing at which hierarchical abstraction levels anomalies manifest across datasets (Figure~\ref{fig:hier_anomaly}), showcasing three level-specific anomalies that \app{} found with LLM-generated explanations in our Appendix~\ref{appendix: exp}, and achieving significant F-1 improvement (82.76\% → 92.83\%) over SOTA
   methods, while reducing LLM usage to  1.1–3.3\% of test sequences.

\end{enumerate}

\begin{figure*}[t]
    \centering
    \includegraphics[width=0.99\linewidth]{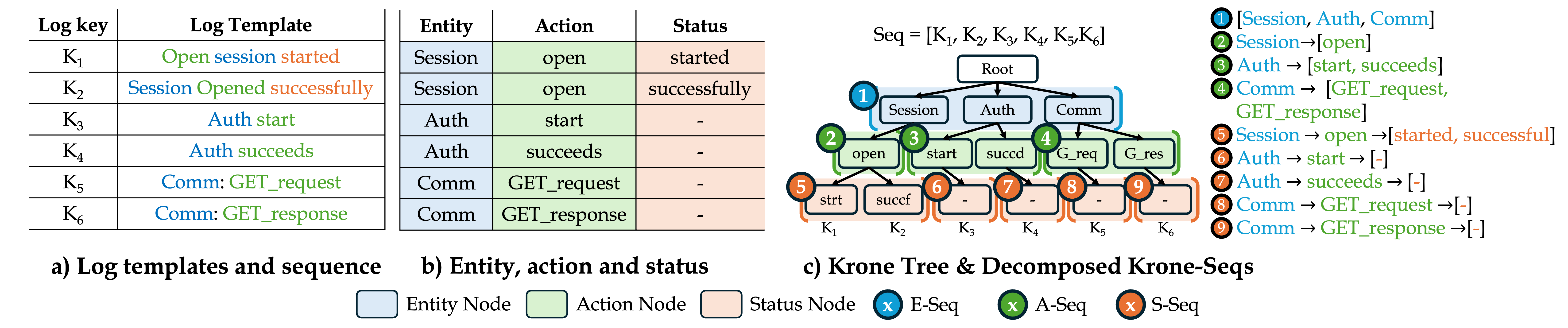}
    \vspace{-10pt}
    \caption{\datamodel{} of Log Data. a) the example log sequence $L$ with log keys and templates. b) the entities, actions, and statuses extracted from the log templates. c) \app{} Tree as the data schema, and \appseqs{} decomposed from example log sequence $L$. }

    \label{fig:data_model}
    \vspace{-12pt}
\end{figure*}

\section{Related Work}
\label{sec:related_work}

In early systems, 
engineers relied on keyword searches and rule matching to identify suspicious 
logs~\cite{10.5555/1024753.1024780,10.5555/1052676.1052694}. However, the growing scale of modern logs and 
sophisticated failures and attacks render such manual approaches too labor-intensive and error-prone~\cite{experiencereport}. 
Leveraging deep learning models, log-based anomaly detection has shifted toward automatically identifying anomalous 
execution behaviors from logs~\cite{Ali_2025}.  Depending on the availability of anomaly labels,  approaches can be 
categorized into supervised and unsupervised methods. 
Supervised approaches~\cite{logrobust,NeutralLog} formulate detection as a binary classification task that requires labeled normal and abnormal sequences. 
Due to the scarcity of labeled anomalies in real-world systems,
 unsupervised methods~\cite{deeplog,OC4SEQ,RT-log,logbert,meng2019loganomaly,cat,pluto,le2021log,logsy} are more widely adopted in practice.
 Most unsupervised methods first learn normal system behavior and then identify anomalies by predicting the next log key.
 A log sequence is flagged as anomalous if the true next log key does not appear among the model’s top-$k$ predictions, with $k$ a tunable hyperparameter.


Recent work has explored 
LLMs for 
tasks
from log parsing to log anomaly detection. LLM-based parsing methods such as LLMParser~\cite{LLMParser} and LibreLog\cite{LibreLog} combine generative LLM reasoning with semantic clustering to achieve strong parsing accuracy.  
Prompt-based frameworks such as LogPrompt~\cite{logprompt}, 
LogGPT~\cite{han2023loggptloganomalydetection}, 
and 
RAG-based frameworks~\cite{zhang2025xraglog,RAGlog} apply LLMs to log anomaly detection. 
These state-of-the-art methods invoke LLMs either on individual log messages or exhaustively over all test log sequences.  
Moreover, as sequences grow longer, beyond their high
computational costs,
LLMs also suffer from the lost-in-the-middle problem~\cite{liu2023lost}, where mid-context information is underutilized during reasoning, risking
to 
compromise detection effectiveness. 
These studies reflect the growing potential
for leveraging LLMs in log analysis, while underscoring the importance of cost-aware and long-context–robust integration.

\color{black}

\section{Preliminaries}
\label{sec:preliminary}


\noindent\textbf{Log Parsing.} The raw log messages are formatted  strings generated by source code, where static textual templates—such as “Received block \verb|<*>| of size \verb|<*>| from \verb|<*>|”—include placeholders (\eg \verb|<*>|) that are filled with variable parameter values such as IP addresses, job IDs, \etc \cite{experiencereport, landauer2023deep, le2022log, le2021log, parsertools, RT-log}. These messages are then continuously appended into log files. Log parsing methods—such as SLCT~\cite{SLCT}, IPLoM~\cite{IPLoM}, LKE~\cite{LKE}, Spell~\cite{SPELL}, and Drain~\cite{drain} aim to convert textual log messages into tabular data by recovering each message’s static template (\eg “Received block \verb|<*>| of size \verb|<*>| from \verb|<*>|”), denoted as $t$, as well as the parameters. Each template is assigned a unique ID (or log key, denoted as $k$), along with a list of extracted variable parameters~\cite{experiencereport}.

\begin{definition}[\textbf{Log Key and Template}]
    \label{def:logkey}
    After parsing, each log message 
    is mapped to its matching template $t$ 
    Each 
     log template $t$ is assigned a unique identifier as log key $k$.

\end{definition}

\noindent\textbf{Log Partitioning.} After parsing, logs are commonly partitioned into \textit{log sequences} $L$ as a standard preprocessing step, using either identifier-based or window-based strategies~\cite{RT-log}.
Identifier-based partitioning groups logs by IDs such as node, domain, or job ID~\cite{RT-log, experiencereport}, but these often reflect routing or instance traces rather than true execution units and depend on naming quality.
Window-based partitioning, widely used by prior work~\cite{deeplog, OC4SEQ, meng2019loganomaly, logrobust, le2022log}, slices logs using count- or time-based windows~\cite{RT-log, experiencereport}.
Both strategies can be used independently or jointly. After parsing and partitioning, logs are represented as sequences of log keys and templates.

\begin{definition}[\textbf{Log Sequence}]
    \label{def:logkey_template_seq}
    With $(k, t)$ representing a 
    log key and template pair, a log sequence is defined as an ordered list: $L = [(k_1, t_1), (k_2, t_2), \dots, (k_{|L|}, t_{|L|})]$, where each $(k_i, t_i)$ corresponds to a log-key template pair 
    observed in sequential order i with $i=1,2, \cdots$
\end{definition}

\section{\app{} Log Modeling}

\label{sec:data_model} 

\noindent\textbf{Hierarchy Abstraction.} 
The \datamodel{} provides an abstraction for hierarchical log modeling, which can be instantiated into a system-specific 
hierarchy tailored to the semantics of a given application. 
This abstraction is grounded in the following
key observation: log messages generally follow a consistent semantic pattern, where a {\it{status}} describes an  {\it{action}} performed on an {\it{entity}}. Figure~\ref{fig:data_model}(a) illustrates an example log sequence 
$L = [(k_1,t_1), (k_2,t_2), \dots, (k_6,t_6)]$ of the log key and templates, as defined \ref{def:logkey_template_seq}. The template $t_1$ for log key $k_1$ “Open session started”, for example, includes a component or process (\eg  $Session$), an action (\eg  $Open$), and an optional status (\eg  $Started$) indicating the outcome. Based on this structure, \datamodel{} defines an abstract three-level schema of the {\it{topics: entities, actions,}} and {\it{statuses. }}
They serve as the basis for instantiating \textit{application-specific log hierarchies}.

\begin{enumerate}[leftmargin=6mm,topsep=0mm,itemsep=0mm]
    \item\textit{Entity.} A key software component or resource involved in an operation, typically expressed as a noun phrase (\eg $Session$).

    \item\textit{Action.} An operation performed by or on an entity, usually described by a verb or verb phrase (\eg $Open$). The set of actions defines the functional behaviors of an entity.

    \item\textit{Status.} The outcome or condition of an action, often expressed as an adjective or noun (\eg $Successful$). The set of statuses defines the possible results of an action.
\end{enumerate}

\noindent\textbf{Hierarchy Extraction.} We frame the extraction of a 
 log hierarchy for a particular application
as a Named Entity Recognition (NER) and Entity Linking (EL) task, with the later identifying and disambiguating entities from text into predefined semantic types~\cite{NER_survey,Entitylinking_survey}. NER techniques span a range of approaches, including rule-based systems, traditional machine learning models, deep learning methods, and more recently LLM-based  techniques~\cite{wang2023gptnernamedentityrecognition, ashok2023promptnerpromptingnamedentity, villena2024llmnerzerofewshotnamedentity}. While conceptually any of 
 NER techniques can be plugged in,
 for example by training a model with a large corpus of annotated data,   inspired by GraphRAG~\cite{graphRAG},  we adopt an LLM-based method to extract domain-specific entities, actions, and statuses directly from log templates  due to its simplicity and ease of deployment.
 
 To illustrate, consider the log key $k_1$ with the template "Open session started." Our approach identifies $Session$ as the entity, $Open$ as the action, and $Started$ as the status—capturing the inherent structure of log data. Figure~\ref{fig:data_model}(b) shows the extracted topics. As illustrated in Figure \ref{fig:data_model}(c), the instantiated hierarchy is then stored as a tree structure, named \textit{\app{} Tree}, which allows for fast retrieval of the three-level topics for a given log key or template.  

\begin{definition}[\textbf{\app{} Tree}]
\label{def:tree}
Let $T = (V, R)$ be a rooted, three-level tree with node set $V$ and parent-child relations $R$.
Let $v^r$ be the root node, $V^E$, $V^A$, and $V^S$ represent the entity, action, and status nodes respectively, where $V = V^E \cup V^A \cup V^S$. The hierarchy is defined by the parent relation:
\[
v.parent =
\begin{cases}
v^r, & \text{if } v \in V^E;\\
v^e \in V^E, & \text{if } v \in V^A;\\
v^a \in V^A, & \text{if } v \in V^S.\\
\end{cases}
\]

Each status node $v \in V^S$ is associated with a log key $k$ and its corresponding log template $t$.
\end{definition}

\noindent\textbf{Hierarchy for Log Sequence Decomposition.} Guided by the \app{} tree,  
given a log sequence, \app{} recursively decomposes it into sequences of entity nodes, action nodes, and status nodes, namely \appseqs{},  at different levels. For higher levels such as \entityseqs{} and \actionseqs{}, they are collapsed by removing consecutive duplicates since we only care about transitions between unique nodes on the current level, \eg transition between unique entities, while the actual  differences in actions of an entity can be captured in the lower level. Meanwhile, we do not collapse  the \statusseqs{} to preserve all status transitions. 

\begin{definition}[\textbf{\app{} Seq}]
    \label{def:seq}
    Given the \app{} tree, $T=(V, R)$, a \appseq{} $S$ is a list of sibling nodes with a common parent.   
    Formally, it can be denoted as:

    \begin{equation}
        S=p\rightarrow[v_1,v_2 \dots v_n]\subseteq V
    \end{equation}
    where $\forall i\in\{1,2\dots n\}$, $v_i.parent=p$ and  $v_i\neq v_{i+1}$. 
    \end{definition}

\appseqs{} can exist at any level of the hierarchy: at the entity (Level 1), action (Level 2), or status (Level 3).
As in Figure~\ref{fig:data_model}(c), the log sequence $L$ can be decomposed into:

\begin{itemize}[leftmargin=5mm,topsep=0mm,itemsep=0mm]
    \item\textit{\entityseq{}.}  At the highest level, the collapsed list, \ie list without consecutive duplicates, of entities of $L$ forms an \entityseq{}, \coloredcircled{entityblue}{1}, represented as $v^r\rightarrow[Session, Auth, Comm]$, which reflects the transition among the entities.

    \item\textit{\actionseq{}.} For each entity in the \entityseq{}, an \actionseq{} captures the sequence of actions under the parent entity. For example, \coloredcircled{actiongreen}{3} shows the \actionseq{} for $Auth$: $Auth$ $\rightarrow$ [$start$, $succeeds$]. By doing this for all entities, the example log sequence $L$ is decomposed into three \actionseqs{}, \coloredcircled{actiongreen}{2}– \coloredcircled{actiongreen}{4}.
    
    \item\textit{\statusseq{}. } Similarly, for each action, a \statusseq{} records its state transitions. E.g., \coloredcircled{statusorange}{5} represents the \statusseq{} for the $open$ action under $Session$: $Session$ $\rightarrow$ $open$ $\rightarrow$ [$started$, $successful$]. By doing this for all action nodes, the example log sequence $L$ is decomposed into four \statusseqs{}, \coloredcircled{statusorange}{5}– \coloredcircled{statusorange}{9}.
\end{itemize}


\noindent\textbf{\appseq{} Properties.} \appseqs{} exhibit three key properties that support effective and efficient detection. (1)
\textbf{\horizontal{} Property:}
They are hierarchically organized  by level (\eg entity, action, status), 
where siblings \appseqs{} have the same parent. This structural locality enables detection within smaller-scoped contextually-coherent groups.
%
(2) \textbf{\atomic{} Property}. The decomposition yields semantically distinct 
\appseqs{} that represent the minimal recurring behavioral units of the system. This is supported by our empirical results (Figure~\ref{fig:cadinality}), which show that the number of \appseqs{} is drastically smaller than the number of full log sequences. 
Together, \horizontal{} and \atomic{} allow \app{} to support adaptive scalable detection by operating on modular self-contained units rather than entire log sequences.
(3)
\textbf{\vertical{} Property}. 
\appseqs{} have inherent cross-level nesting.
This arises from the underlying tree structure of the nodes: each \appseq{} is a list of nodes at a specific level, and when a node within a higher-level \appseq{} is replaced by its corresponding lower-level \appseq{}, a nested structure emerges.

\vertical{} introduces different representational granularities for \appseqs{}, which we define as the \localformat{} format, ignoring lower-level details, and the \nestedformat{} format, recursively incorporating them.

\begin{definition}[\textbf{\localformat{} vs. Nested-Aware Format of \appseqs{}}]
    \label{def:format}
    Given a hierarchy level $l \in \{0,1,2\}$ where $0,1,2$ stand for status, action, and entity, respectively, and a \appseq{} $S^l$ at level $l$, we define two formats of \( S^l \) based on whether lower-level semantics are incorporated:
    \begin{itemize}[leftmargin=5mm,topsep=0mm,itemsep=0mm]
        \item\textit{\localformat{} Format:} A \appseq{} is represented homogeneously  by its nodes on the current level. That is,
        \begin{equation}
            S^l = [v^l_1, v^l_2, \dots, v^l_n]\in V^l
        \end{equation}
        \item\textit{\nestedformat{} Format:} The \appseq{} is recursively expanded by replacing each node $v_i^l$ with its  lower-level \appseq{} $\tilde{S}^{l-1}$, yielding the nested-aware format $\tilde{S}^l$:    
        \begin{equation}
            \tilde{S}^{l} = 
            \begin{cases}
                S^{l} & \text{if } l = 0 \\
                [\tilde{S}^{l-1}_1, \tilde{S}^{l-1}_2, \dots, \tilde{S}^{l-1}_n] & \text{if } l \geq 1 \\
            \end{cases}
        \end{equation}
    \end{itemize}

    This format preserves the 
     semantics by recursively embedding lower-level behavior into higher-level representations.
\end{definition}

These two formats reflect a trade-off between efficiency and semantic richness. The \localformat{} format enables fast detection by analyzing high-level patterns in isolation, while the \nestedformat{} format offers greater accuracy by incorporating the hierarchical context yet  with an increase in
 computational costs.

\section{Model Instantiation and Utilization}
\label{sec:model_construct}

Since log templates already encode the semantic types of logs and form a much smaller set than raw log messages, we extract entities, actions, and statuses directly from templates to construct the \app{} Tree—avoiding unnecessary processing of massive log files. The inferred hierarchy is then used to decompose each log sequence into multi-level \appseqs{}.

\noindent\textbf{Topic Extraction.}
Given each log template, we use in-context LLM learning with manually-made  examples 
to extract the entity, action, and status, where the prompts can be found  here\footnote{\url{https://github.com/LeiMa0324/Krone_official/blob/main/tree_extraction/EXTRACT_PROMPTS.py}\label{extrac_prompt}}.
The extraction prompt asks the LLM to consider the relationship between the entity, action, and status, \ie an action is extracted based on the extracted entity, and the status is extracted for the action.  
To enforce structural consistency, we perform an additional refinement pass. 
For each template, we prompt the LLM to select or generate the most appropriate entity from the pool of extracted entities. 
Conditioned on the selected entity, we then prompt the LLM to select or generate the most suitable action from the set of extracted actions associated with that entity. 
Since each template is associated with a unique status by construction, we do not perform refinement for the status level.

\noindent\textbf{\app{} Tree Construction.} After extracting entities, actions, and statuses from templates, we build the \app{} Tree by structuring them into a three-level hierarchy. A single root connects to all unique entities; each entity links to the distinct actions associated with it; and each action connects to its corresponding statuses found in templates. To preserve a one-to-one mapping, each status is uniquely tied to its log key and template. This process efficiently organizes all log keys and templates under a semantic hierarchy.

\noindent\textbf{Top-Down Sequence Decomposition.} While Figure~\ref{fig:data_model}(c) provides an intuitive illustration of the recursive decomposition process, the process is now formally defined by Algorithm~\ref{algo:decompose}. 
To support this recursive decomposition, we propose a utility algorithm \textsc{GenerateKroneSeq}. Given a log sequence and a specified level, it returns a \appseq{} by mapping each log key to a node on the specified level in the \app{} Tree. It also returns the corresponding list of log chunks $\mathcal{X}$. This utility algorithm is invoked at each level to perform segmentation with time complexity linear in the sequence length. Due to the limited space, we provide the pseudo code in the Appendix~\ref{appendix: method}.

Given a log sequence and the \app{} tree, Algorithm~\ref{algo:decompose}  generates (1) a \entityseq{} $S_E$, (2) a list of \actionseqs{} $\mathcal{S_A}$ for each entity node $v^e \in S_E$ and (3) a list of \statusseqs{} $\mathcal{S_S}$ for each action node $v^a \in S_A \in \mathcal{S_A}$. Line 2 first decomposes $L$ in the entity level to generate the \entityseq{} $S_E$ and the chunks $\mathcal{X}_E$ for each entity node in $S_E$. Then, in Line 4, for each chunk $x_e \in \mathcal{X}_E$, Line 5 calls the utility again to further decompose it in the action level, generating the \actionseq{} $S_A$ for each entity node and the chunk list $\mathcal{X}_A$. Lines 6-7 maintain the details of the \actionseq{} $S_A$ and append it to the list of \actionseqs{} $\mathcal{S_A}$. Similarly, for each chunk $x_a\in \mathcal{X}_A$, Lines 8-11 repeat this process to generate the \statusseq{} $S_S$, maintain its details, and add it into the list of \statusseqs{} $\mathcal{S_S}$. At last, Line 12 returns all \appseqs{} as output. Table \ref{tab:tab_decompose} shows the results
of the decomposed \appseqs{}.

\begin{table}[t]
    \caption{Details of \appseqs{}. Log chunks of templates are omitted for simplicity. }
    \vspace{-5pt}
    \centering
    \resizebox{1.0\linewidth}{!}{
    \begin{tabular}{c|c|c|c|c}
        \toprule
        \textbf{Index} & \textbf{Parent} & \textbf{\appseq{}} & \textbf{Log Chunk} & \textbf{Nested Sem} \\
        \hline
        \coloredcircled{entityblue}{1} & root & \textcolor{entityblue}{[Session, Auth, Comm]} & $[k_1, k_2, k_3, k_4, k_5, k_6]$ & [\coloredcircled{actiongreen}{2}, \coloredcircled{actiongreen}{3}, \coloredcircled{actiongreen}{4}] \\
        \hline
        \coloredcircled{actiongreen}{2} & \textcolor{entityblue}{Session} & \textcolor{actiongreen}{[open]} & $[k_1, k_2]$ & \coloredcircled{statusorange}{5} \\
        \hline
        \coloredcircled{actiongreen}{3} & \textcolor{entityblue}{Auth} & \textcolor{actiongreen}{[start, succd]} & $[k_3, k_4]$ & [\coloredcircled{statusorange}{6}, \coloredcircled{statusorange}{7}] \\
        \hline
        \coloredcircled{actiongreen}{4} & \textcolor{entityblue}{Comm} & \textcolor{actiongreen}{[GET\_req, GET\_res]} & $[k_5, k_6]$ & [\coloredcircled{statusorange}{8}, \coloredcircled{statusorange}{9}] \\
        \hline
        \coloredcircled{statusorange}{5} & \textcolor{entityblue}{Session} → \textcolor{actiongreen}{open} & \textcolor{statusorange}{[started, succf]} & $[k_1, k_2]$ & - \\
        \hline
        \coloredcircled{statusorange}{6} & \textcolor{entityblue}{Auth} → \textcolor{actiongreen}{start} & [-] & $[k_3]$ & - \\
        \hline
        \coloredcircled{statusorange}{7} & \textcolor{entityblue}{Auth} → \textcolor{actiongreen}{succd} & [-] & $[k_4]$ & - \\
        \hline
        \coloredcircled{statusorange}{8} & \textcolor{entityblue}{Comm} → \textcolor{actiongreen}{GET\_req} & [-] & $[k_5]$ & - \\
        \hline
        \coloredcircled{statusorange}{9} & \textcolor{entityblue}{Comm} → \textcolor{actiongreen}{GET\_res} & [-] & $[k_6]$ & - \\
        \hline
    \end{tabular}}
    \label{tab:tab_decompose}
\end{table}


\begin{algorithm}[t]
\caption{\textsc{\algo{}}}
\label{algo:decompose}
\begin{algorithmic}[1]
\Require Log sequence $L=[k_i]$, \app{} tree $T=(V,R)$ with root $v_r$.
\Ensure \entityseq{} $S_E$, list of \actionseqs{} $\mathcal{S_A}$, list of \statusseqs{} $\mathcal{S_S}$.
\State $S_E \gets [\ ],\quad \mathcal{S_A} \gets [\ ],\quad \mathcal{S_S} \gets [\ ]$
\State $\mathcal{X}_E, S_E \gets \Call{\textbf{\partitionalgo}}{L, T, \text{entity}}$
\State $\Call{MaintainDetail}{S_E, L, v_r, \text{null}}$ \Comment{chunk, parent, nest}
\For{$(x_E, v_e)$ \textbf{in} $\Call{Zip}{\mathcal{X}_E, S_E}$}
    \State $\mathcal{X}_A, S_A \gets \Call{\textbf{\partitionalgo}}{x_E, T, \text{action}}$
    \State $\Call{MaintainDetail}{S_A, x_E, v_e, S_E}$
    \State $\Call{ListAppend}{\mathcal{S_A}, S_A}$
    \For{$(x_A, v_a)$ \textbf{in} $\Call{Zip}{\mathcal{X}_A, S_A}$}
        \State $\mathcal{X}_S, S_S \gets \Call{\textbf{\partitionalgo}}{x_A, T, \text{status}}$
        \State $\Call{MaintainDetail}{S_S, x_A, v_a, S_A}$
        \State $\Call{ListAppend}{\mathcal{S_S}, S_S}$
    \EndFor
\EndFor
\State \Return $S_E, \mathcal{S_A}, \mathcal{S_S}$
\end{algorithmic}
\end{algorithm}




\noindent\textbf{Complexity Analysis. } We analyze the end-to-end complexity of hierarchy extraction, tree construction, and sequence decomposition. 
Let $|T|$ denote the number of unique templates. 
Topic extraction processes each template with a constant number of operations and thus runs in $O(|T|)$ time. 
Since the size of \app{} Tree is bounded by $O(3|T|)$, its construction, implemented using a hashmap-based node index, likewise runs in $O(|T|)$ time.  For the decomposition in Algorithm~\ref{algo:decompose}, 
let $|L|$ denote the length of the input log sequence. 
Line 2 runs in $O(|L|)$ time. 
In the for-loop (Line 4), each chunk $x_E$ is processed in $O(|x_E|)$ time (Line 5), and since the chunk lengths sum to $|L|$, the total cost of this loop is $O(|L|)$. 
The same  applies to Line 9, yielding another $O(|L|)$ term. 
All remaining operations (Lines 6--7 and 10--11) are linear in the sequence length. 
Thus, the overall time complexity of Algorithm~\ref{algo:decompose} is $O(|L|)$. Let $n$ be the number of all sequences, and $|L|_{max}$ be the max of the length of the sequences, then the overall end-to-end complexity is bounded by $O(n|L|_{max}+|T|)$.

\section{Modular \appseq{}  Detection Principles}
\label{sec:modular_detection}

By decomposing a log sequence into \appseqs{}, the detection task is transformed into more manageable modular subtasks at the \appseq{} level. In this section, we introduce the principles of modular \appseq{}-level detection with two detection strategies based on two complementary formats of \appseqs{}—\localformat{} and \nestedformat{} (Definition~\ref{def:format}). We later implement these strategies with concrete detectors and adaptively combine them to balance efficiency and semantic richness in Section~\ref{sec:detection_framework}.

 To this end, we now present the modular \appseq{} detection principles with two complementary execution strategies.

\begin{figure*}[h]
    \centering
    \includegraphics[width = 0.99\linewidth]{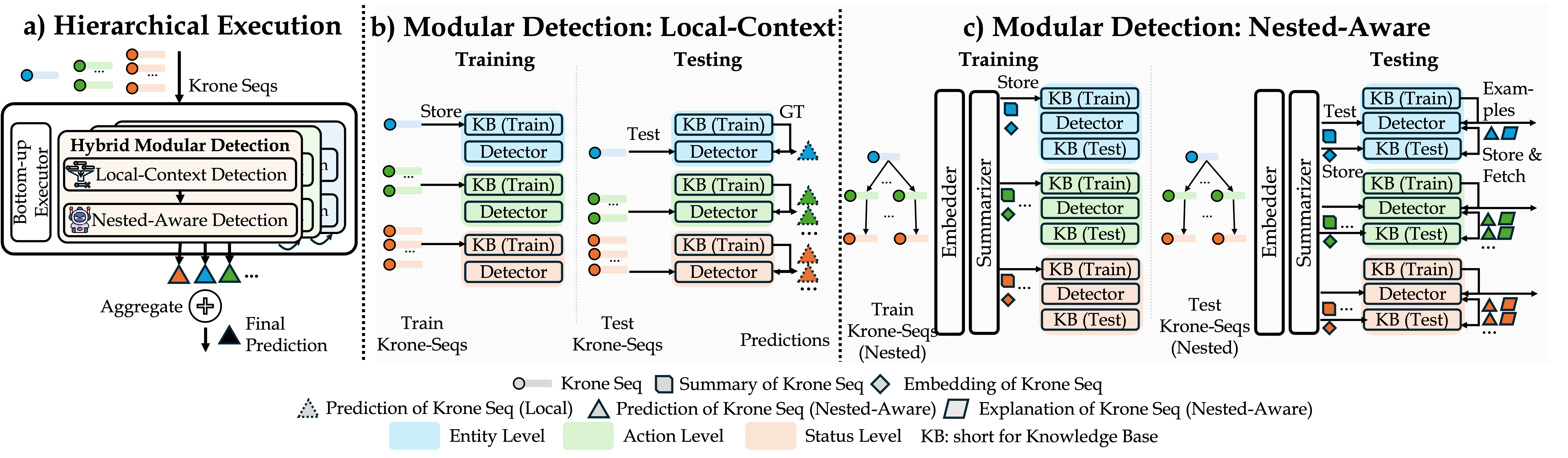}
    \vspace{-10pt}

    \caption{\app{} Framework. a) Hierarchical Bottom-up Execution. b) Modular \appseq{} Detection: \localformat{} Strategy with pattern matching. c) Modular \appseq{} Detection: \nestedformat{} Strategy with LLM. }
    \label{fig:arch}
    \vspace{-10pt}
\end{figure*}


\noindent\textbf{\appseq{} Detection.} Through its hierarchical design, \app{} decomposes normal training log sequences into \textit{normal training \appseqs{}} that serve as ground truth for \appseq{}-level detection. 

\textbf{\textit{Training Knowledge Base.}} To support this process, a training knowledge base is constructed for each node in the hierarchy, storing
the
normal \appseqs{}
under it. 
For instance, the knowledge base of the root node contains all normal \entityseqs{}, while each entity and action node maintains the normal \actionseqs{} and \statusseqs{} associated with it, respectively.  As in Table \ref{tab:tab_decompose}, we store the \appseqs{}, together with their parents, log chunks, and \vertical{}.

\textbf{\textit{Detection Procedure.} } Through hierarchical decomposition, a test log sequence is similarly divided into test \appseqs{}, each evaluated against its normal counterparts in the corresponding training knowledge base. For instance, a test \appseq{} under its parent will be verified against its normal siblings \appseqs{}.  Since this detection process adheres to a standard matching-and-verification paradigm, \app{} flexibly accommodates a wide range of detectors—whether operating on symbolic patterns (\eg chunks of log keys or node sequences) or on semantic representations (e.g., chunks of log templates)—which can be seamlessly integrated 
into our system for \appseq{}-level detection.

\noindent\textbf{\appseq{} Detection Strategies. } While the individual \appseq{}-level detection follows the standard procedure, the two 
%
formats
\ie \localformat{} and \nestedformat{} 
 lead to different detection strategies.

\textbf{\textit{\localformat{} Strategy.}}
By disregarding the \vertical{}, \appseqs{} in the \localformat{} format correspond directly to the nodes at the current level.
This design allows \app{} to treat each \appseq{} as an independent subproblem, so that they can be executed in parallel, thereby enhancing computational efficiency.

\textbf{\textit{\nestedformat{}  Strategy.}}
While local detection efficiently identifies irregularities within a single level, higher-level \appseqs{} are inherently compositions of lower-level \appseqs{}.
However, as we recursively decompose the log sequence into nested \appseqs{} in
a top-down manner (Section \ref{sec:model_construct}), 
this structure naturally motivates a bottom-up detection strategy: higher-level detection is more expensive and depends on the lower-level results. Starting from status level,  predictions and pattern information from child \appseqs{} are propagated upward to guide the detection of their parent \appseqs{}.
By recursively executing along the hierarchy, this \nestedformat{} format represents  \appseqs{} 
the system behavior
in a lossless manner 
—especially beneficial for detecting complex context-dependent anomalies.

While both strategies demonstrate the adaptability of \app{}’s hierarchy, they expose a clear trade-off between efficiency and accuracy.
The \localformat{} strategy scales efficiently but may overlook cross-level semantics, whereas the \nestedformat{} strategy captures semantics at the cost of dependent execution.

To integrate these complementary strengths, we next introduce a unified \app{} Detection Framework that realizes these principles and combining them into hybrid modular detection with adaptive routing between detectors, and a resource-optimized hierarchical execution that efficiently schedules the modular detection subtasks.

\section{Hierarchical \app{} Framework: Execution and Optimization}

\label{sec:detection_framework}

Here,
we realize the modular detection principles introduced in Section~\ref{sec:modular_detection}
by incorporating them 
into a hybrid modular detection that dynamically routes between a local  pattern-matching detector for efficient symbolic analysis (Figure \ref{fig:arch}(b)) and a nested LLM-based detector for semantic reasoning (Figure \ref{fig:arch}(c)). 
Based on the hybrid modular detection, we propose the hierarchical \app{} detection framework (Figure \ref{fig:arch}(a)), which executes modular detection tasks bottom-up with resource-aware optimizations—enabling cost-efficient anomaly detection over large-scale log data.

\vspace{0.3em}
\noindent\textbf{\localformat{} Detector via Pattern Matching.}  
The decomposition not only reduces the size of the detection task  but also divides it into a set of more tractable subproblems. 
As a result, 
detection at the
\appseq{}-level 
become substantially easier than sequence-level detection,
 making even simple detectors including exact pattern matching  highly effective.
Our empirical results (Table~\ref{tab:comapre_PM}) further validate this observation.

As in Figure \ref{fig:arch}(b), normal \appseqs{} are firstly stored into the knowledge bases during training as ground-truth, and during testing, each test \appseq{} is evaluated in its \localformat{} representation as a node sequence and flagged as anomalous by this detector if it does not appear in the corresponding training knowledge base. This strict detector quickly passes through confident normal patterns. 

\vspace{0.3em}
\noindent\textbf{\nestedformat{} Detector via LLM Reasoning.}  Beyond enabling simple detectors on complex data, decomposition also makes large-scale LLM-based anomaly detection feasible by operating on \appseqs{} instead of full log sequences.
Moreover, 
due to the atomicity and reusability of \appseqs{}, each unique \appseq{} needs to be processed only once.
this yields resource savings up to $43.7\times$ as confirmed by our empirical evaluation (Section~\ref{sec:exp:hier_viz}, Figure~\ref{fig:tokens}). 

To this end, the \app{} hierarchy naturally supports LLM-based detection. As illustrated in Figure~\ref{fig:arch}(b), each \appseq{} is embedded (of the log key chunk) and summarized into a textual description during both training and testing, and these artifacts are stored in level-specific knowledge bases. During testing, following the standard in-context learning (ICL) paradigm~\cite{brown2020language,min2022rethinking}, the summary of a test \appseq{}, along with the summaries of its top-$k$ most similar normal siblings retrieved from the knowledge bases as examples, are formatted into a prompt to the LLM for the prediction and an explanation. To avoid redundant re-summarization and ensure semantic consistency, we leverage the nested structure and summarize in a bottom-up manner: each lower-level \appseq{} is first summarized by an LLM. These summaries are recursively used as building blocks to compose the summary of  a higher-level \appseq{} using LLM. The prompts used in our method are publicly available\footnote{\url{https://github.com/LeiMa0324/Krone_official/blob/main/krone_hierarchy/PROMPTS.py}}. 

While our focus is on enabling a scalable LLM detection pipeline, the framework itself is detector-agnostic: more advanced LLM reasoning strategies (e.g., chain-of-thought~\cite{wei2023chainofthoughtpromptingelicitsreasoning}, ReAct~\cite{react}) can be seamlessly integrated.

\noindent\textbf{Hierarchical Bottom-up Execution.} Figure \ref{fig:arch}(a) illustrate the overall hierarchical execution framework of the modular detection tasks. Given a test log sequence and its decomposed test \appseqs{}, the modular detection produce a prediction for each test \appseq{}, and the predictions are aggregated into the final prediction for the log sequence.

$\bullet~$\textit{Hybrid Modular \appseq{}-level Detection.} 
For \appseq{}-level detection, \app{} dynamically routes execution between 
calling either of the two detectors to balance efficiency and effectiveness. Each test \appseq{} is first processed by the pattern-matching detector for  strict filtering—quickly discarding known normal \appseqs{}. 
Only those that do not match any known pattern are then delegated to the LLM for semantically enriched reasoning. Guided by retrieved demonstration examples, the LLM generates both the anomaly prediction and the explanation.

$\bullet~$\textit{Test Knowledge Base.} To fully exploit the atomicity of \appseqs{}, \app{} maintains a test knowledge base for each level, mirroring the training knowledge bases. For every test \appseq{}, its artifacts, including the LLM-generated prediction, explanation, and summary, as well as the embedding, are stored for reuse. When the same test \appseq{} reappears (which is confirmed by our experimental evaluation in Figure \ref{fig:reusability}), these artifacts are directly retrieved by matching its log chunk—eliminating redundant LLM queries.
    
$\bullet~$\textit{Early-Exit Optimization Strategy.} Following the bottom-up execution paradigm, \app{} first evaluates all \statusseqs{} of a test sequence, then progresses to \actionseqs{} and \entityseqs{}. To avoid unnecessary computation on higher-level \appseqs{} with longer context, \app{} employs an early-exit rule: once any \appseq{} is detected as abnormal, execution stops and the entire log sequence is flagged anomalous. This reflects the intuition that a single abnormal segment is sufficient for sequence-level detection.

\section{Experimental Setup}

\subsection{Datasets}
\label{sec:dataset}
We evaluate \app{} on four real-world log datasets from large-scale high-performance computing systems. This include three public benchmark datasets (BGL, HDFS, ThunderBird), commonly adopted in prior work~\cite{logbert, deeplog,RT-log,cat,OC4SEQ,pluto} and a proprietary industry-strength dataset from ByteDance Cloud (Bytedance-IaaS). Details are provided in Table~\ref{tab:datasets}.
\begin{itemize}[leftmargin=5mm,topsep=0mm,itemsep=0mm]

    \item\textbf{BlueGene/L (BGL)} \cite{oliner2007supercomputers} collected from the BlueGene/L supercomputer system at Lawrence Livermore National Laboratory (LLNL) in Livermore, California. Each log entry includes an alert label, a timestamp, and log content. The dataset comprises 4,747,936 log messages, with 348,460 classified as alerts.

    \item\textbf{Hadoop Distributed File System (HDFS)} \cite{HDFS} gathered from a Hadoop-based MapReduce cloud environment using benchmark queries. Each log consists of a timestamp, log level, content linked to one or multiple block IDs. The dataset contains 11,175,629 log messages, segmented into sequences based on block IDs. Experts labeled each sequence as normal/abnormal.

    \item\textbf{ThunderBird } \cite{oliner2007supercomputers} collected from the Thunderbird supercomputer system at Sandia National Laboratories (SBNL) in Albuquerque. Logs are generated on local nodes and aggregated by a log server. Each log entry includes an event with an alert label, a timestamp, a source node ID, and log content. The raw dataset consists of 211,212,192 logs. For training efficiency, we apply subset sampling following \cite{cat,pluto}, selecting the first 2 million log messages, of which 110,232 are labeled as alerts.

    \item\textbf{Bytedance-IaaS (Industry)} from the Infrastructure as a Service  (IaaS) layer of  Bytedance Cloud Service. Each log entry includes a timestamp, flow ID, flow Status, and content. The logs can be grouped by flow ID, with each flow representing a sequence of logs generated by a customer’s request for a specific Action (\eg, creating or deleting a virtual machine). Each flow is assigned a Status, indicating if the flow is executed successfully. Abnormal flows are rolled back automatically or manually by engineers. Successful flows are normal sequences, while rollback flows are considered to be  abnormal. The dataset contains a total of 1,053,747 logs.
\end{itemize}

\begin{table}[t]
\caption{Dataset Statistics. }
\vspace{-5pt}
\centering
\resizebox{0.95\linewidth}{!}{
\begin{tabular}{l|ccc|c}
    \toprule
    \textbf{Dataset}  & \textbf{\#Sequence} & \textbf{\#Unique Log keys} & \textbf{Anomaly (\%)} & \textbf{Train:Test} \\\hline

    HDFS & 591,899 & 48& 2.8& 1:99 \\
    BGL & 231,564 & 1000&9.5& 2:8\\
    ThunderBird & 99,026 & 1205& 0.48& 2:8\\
    Bytedance-IaaS &  1800& 157 & 16.67& 2:8\\

    \bottomrule
    
\end{tabular}}

\label{tab:datasets}
\vspace{-15pt}
\end{table}

\noindent\textbf{Log Data Pre-processing}. We follow standard log pre-processing~\cite{cat,OC4SEQ,logbert,pluto}. Raw logs are parsed using Drain~\cite{drain} to extract log keys and templates and form log sequences.
For HDFS and Bytedance-IaaS, logs are grouped by block ID and flow ID, respectively. ThunderBird and BGL lack explicit sequence boundaries; we group ThunderBird logs by node ID and segment both datasets using sliding windows.
Sequence labels are provided for HDFS and Bytedance-IaaS. For BGL and ThunderBird, we follow prior work~\cite{logbert,deeplog,pluto,logrobust,meng2019loganomaly}, labeling a sequence as abnormal if it contains at least one abnormal key.
Table~\ref{tab:datasets} reports dataset statistics after preprocessing.

%
\noindent\textbf{Train-Test Split.} Following prior work~\cite{deeplog,logbert,logrobust,meng2019loganomaly,OC4SEQ}, we adopt the standard one-class setting, training only on normal sequences.
Through the \datamodel{}, \app{} improves data efficiency by learning normal patterns from fine to coarse granularity.
Accordingly, we use a 2:8 train–test split for BGL, ThunderBird, and Bytedance-IaaS, and a 1:99 split for the larger HDFS dataset.

\subsection{Comparative Methods}
We compare \app{} with the following three groups of state-of-the-art log anomaly detection methods.



\noindent\textbf{Classical Mining 
Methods.} This group includes methods using pattern mining, similarity comparison, and classical machine learning to detect anomalies, including \textit{PCA} \cite{PCA}, \textit{OCSVM} \cite{OCSVM}, and \textit{IsolationForest} \cite{iForest}.

\begin{table*}[t]
    \centering
    \caption{Performance of all anomaly detection methods. Bold for F-1 denotes  winner. }
    \vspace{-5pt}
    \label{tab: main_perform}
    \resizebox{0.8\linewidth}{!}{
    \begin{tabular}{l|ccc|ccc|ccc|ccc}
    \toprule
    Dataset & \multicolumn{3}{c|}{HDFS} & 
    \multicolumn{3}{c|}{BGL} &
    \multicolumn{3}{c|}{ThunderBird} &
    \multicolumn{3}{c}{IaaS (Industry)}\\
    
\midrule
Metric & P (\%) & R (\%) & F-1 (\%)  & P (\%) & R (\%) & F-1 (\%)   & 
P (\%) & R (\%) & F-1 (\%) &P (\%) & R (\%) & F-1 (\%)  \\

\midrule
\midrule
PCA & 97.59&66.52 &79.11  &41.13&23.55&29.95&28.53&99.58&44.35&
33.74&100.00&50.46
\\
OCSVM &
21.21&82.48&33.74&
20.00&18.33&19.13&
9.89&100.00&17.99&
70.59&100.00&82.76\\
IsolationForest &
4.93&31.62&8.54&
21.74&13.13&16.37&
1.72&100.00&3.40&
64.51&100.00&78.43\\
\hline

OC4Seq &2.95 &100.00 &5.75 &11.89 &100.00 &21.25&1.02&100.00&2.02&
20.83&100.00&34.48
\\

LogBert&81.09&65.02&72.17&88.98&97.76&93.16&9.81&100.00&17.87& 25.91&100.00 & 41.16\\
Logsy&95.01&61.04&74.33
&76.06
&99.29
&86.14
&50.47&99.79&67.04& 37.44&99.03 & 54.34\\
DeepLog&94.44&33.41&49.36&79.76&99.22&88.43&49.90&99.79&66.53&68.49 & 100.00& 81.30\\
LogAnomaly&97.12&68.95&80.64&33.81&99.67&50.50&2.01&99.79&3.95&20.83 &100.00 & 34.48\\
LogRobust&95.82&64.69&77.24&33.81&99.67&50.50&2.01&99.79&3.95&20.83 & 100.00& 34.48\\

\midrule

LogPrompt &2.90 &80.92 &5.61 & 11.83 & 99.99& 21.16&4.80&100.00&9.16&
28.57&100.00&16.66\\

\midrule
\textbf{\app} &80.72&99.31&\textbf{89.06}&
92.84 &99.97 &\textbf{96.27}&
60.12&100.00&\textbf{75.10}&
87.13 & 99.33&\textbf{92.83}
\\

\bottomrule    
   \end{tabular}}
   \vspace{-15pt}
\end{table*}

\noindent\textbf{Deep-learning-based Methods.} These methods leverage deep learning techniques for log sequence anomaly detection, including the LSTM/RNN-based methods (\textit{Deeplog}~\cite{deeplog}, \textit{OC4SEQ}~\cite{OC4SEQ}, \textit{LogAnomaly} \cite{meng2019loganomaly},  and \textit{LogRobust} \cite{logrobust}), and the Transformer-based methods (\textit{LogBert}~\cite{logbert},  Logsy\cite{logsy}).

\noindent\textbf{LLM-based Methods.} 
\textit{LogPrompt} \cite{logprompt} is an LLM-based anomaly detection method 
 applied to each single log message.
 Given an individual log message and  hand-coded
 detection rules, it prompts an LLM to generate the detection result and explanation. \textit{XRAGLog}~\cite{zhang2025xraglog} is a RAG-based log anomaly detection method that retrieves normal log sequences as contextual evidence for detecting abnormal sequences and exhaustively applies LLM reasoning to all test inputs. Due to its high costs, we evaluate XRAGLog and \app{} separately in Table~\ref{tab:LLM-comparison}, on a 10\% subsampled dataset of the public benchmark datasets (HDFS-mini, BGL-mini and ThunderBird-mini) with a 5:5 train–test split.

\noindent\textbf{Implementation Details.} We use  public implementations for all comparative methods~\cite{loglizer,deeploglizer,oc4seqcode,logpromptcode,logbertcode}. \textit{LogBERT} uses four Transformer layers (four heads, hidden size 256) with learning rates of $10^{-3}$ for BGL, ThunderBird, and IaaS, and $10^{-4}$ for HDFS, trained for 200 epochs. \textit{DeepLog}, \textit{LogAnomaly}, and \textit{LogRobust} use two-layer LSTMs trained for 100 epochs, with the latter two incorporating TF–IDF semantic features~\cite{deeploglizer}. We use exact pattern matching as the \localformat{} detector for HDFS and IaaS; for BGL and ThunderBird, pattern matching is used at the Status and Action levels, and an automaton is used at the Entity level. We use ChatGPT-3.5 for all LLM-based methods, including \app{}, \textit{LogPrompt}, and \textit{XRAGLog}. \app{} uses $m=5$ as the default number of retrieved normal examples, and outputs direct binary decisions without thresholding.

\noindent\textbf{Evaluation Metrics.} The LLM-detection methods\cite{logprompt,zhang2025xraglog}, including \app{}, directly output binary predictions without producing anomaly scores, which makes the score-based metrics such as AUC-ROC not applicable. Thus, following prior works~\cite{deeplog,logbert,meng2019loganomaly,logrobust,pluto, Bridging_the_Gap, zhang2025xraglog}, we evaluate sequence-level predictions using \textit{Precision}, \textit{Recall}, and \textit{F1-score}. 
Methods like \textit{LogBERT}, \textit{DeepLog}, \textit{LogAnomaly}, and \textit{LogRobust} rely on next or masked log key prediction, flagging a sequence as abnormal if a predicted log key is missing from the model’s top-$k$ predictions (with $k=9$, as recommended by \textit{DeepLog}).
Following the same mechanism, the template-level predictions of \logprompt{} are converted to a sequence-level prediction. For anomaly score or distance-based methods (\eg \textit{PCA}, \textit{IsolationForest}, \textit{OCSVM}, \textit{OC4Seq}), we give them extra advantage, 
reporting the best F1-score  
by sweeping the detection threshold over 20 evenly spaced intervals.

\section{Experimental Evaluation}
\label{sec:exp}

We conduct a comparative study
with 
all methods in Section~\ref{sec:exp:overall}. Studying  our own method \app{}, we analyze its hierarchy in Section~\ref{sec:exp:hier_viz}, conduct an ablation study in Section~  \ref{sec:exp:Ablation},  and evaluate its  LLM effectiveness and costs in Section~\ref{sec:exp:llm_effective}. We provide more experimental evaluations in the Appendix~\ref{appendix: exp}, including F-1 versus different training percentages, the detailed \app{} decomposition results, real-world anomaly examples, and lantency analysis of \app{}.

\subsection{Performance Study of Comparative Methods}
\label{sec:exp:overall}

Table~\ref{tab: main_perform} summarizes the performance of all methods across four datasets. Under a scarce training scenario, \app{} consistently outperforms existing methods due to its hierarchical log-specific learning and detection strategy. While most baselines achieve high recall, they suffer from low precision and high false alarms—indicating they are struggling with learning normal patterns  even for models like \logbert{} that use powerful architectures such as Transformers. In contrast, \app{}’s hierarchical structure enables progressive pattern learning from fine- to coarse-grained levels, while its integration of LLM-capabilities enhances its semantic understanding and generalization. Further, unlike traditional methods that rely on top-$k$ predictions or threshold tuning for best F-1 scores, \app{} produces direct anomaly predictions via exact pattern matching and LLM-based detection. While pattern matching is conservative and prioritizes recall by filtering known normal \appseqs{}, \app{} invokes LLM-based reasoning only for uncertain cases.
By enabling semantic understanding of normal execution behaviors, the LLM component increases precision by removing false positives that pass conservative screening, as further verified in Figure~\ref{fig:demo_num}. 
This conservative screening constrains when LLM reasoning is applied, balancing the precision–recall tradeoff while minimizing unnecessary LLM usage and maintaining high efficiency.

For LLM-based methods, \logprompt{} in Table \ref{tab: main_perform} has the lowest precision due to message-level classification without sequence context, leading to substantial over-detection. XRAGLog in Table \ref{tab:LLM-comparison} also tends to over-detect anomalies even with normal sequences as context, reflecting a limited understanding of long normal execution patterns. It requires one LLM call per test sequence, severely limiting scalability. In contrast, \app{} improves F-1 by hierarchically decomposing sequences and combining pattern matching with selective LLM reasoning, invoking LLMs only when necessary and reducing LLM usage by up to 55.07$\times$ compared to XRAGLog. These results highlight that LLMs must be used judiciously for effective and scalable log anomaly detection.

\color{black}

 \begin{table}[t]
      \caption{Comparison between XRAGLog and \app{}. }
     \label{tab:LLM-comparison}
     \centering
     \resizebox{\linewidth}{!}{
     \begin{tabular}{l|c|cccc}

     \toprule
     Dataset & Method & P(\%)$\uparrow$&R(\%)$\uparrow$&F-1(\%)$\uparrow$&LLM calls$\downarrow$\\\hline\hline
        \multirow{2}{*}{HDFS-mini}& XRAGLog  &50.00 &  65.89 & 56.85& 27,316\\
        & \textbf{\app{}}  &\textbf{93.02}& \textbf{98.92} &\textbf{95.88} & \textbf{496} (55.07$\times\downarrow$)\\\midrule
        \multirow{2}{*}{BGL-mini} & XRAGLog&32.71 & 88.48 & 47.76& 12,096\\
        & \textbf{\app{}} & \textbf{82.20}& \textbf{100.0}& \textbf{90.23} & \textbf{459} (26.3$\times\downarrow$)\\\midrule
    \multirow{2}{*}{ThunderBird-mini} & XRAGLog& 9.05& 92.00 & 16.48& 2,803\\
    & \textbf{\app{}} & \textbf{37.88}& \textbf{100.00}& \textbf{54.95} & \textbf{103} (27.2$\times\downarrow$)\\
        
        \bottomrule
     \end{tabular}}
\vspace{-10pt}
 \end{table}
\color{black}



\begin{figure}[h]
    \centering
    \includegraphics[width=0.98\linewidth]{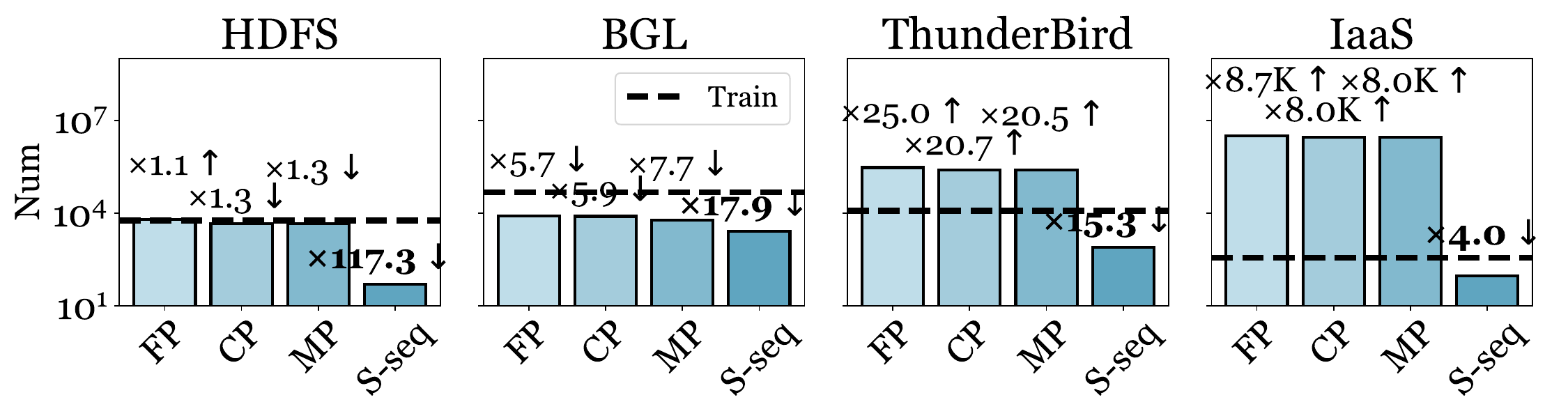}
    \vspace{-10pt}
    \caption{Cardinality reduction achieved by \app{}. Y-axis  in log scale. Black dashed line: total training size; FP: \# frequent patterns. CP: \# closed patterns. MP: \# maximal patterns. S-seq: \#  \statusseqs{}. $\uparrow$: size increase vs. total train size, $\downarrow$: size decrease vs. total train size. 
    }
    \label{fig:cadinality}
    \vspace{-10pt}
\end{figure}

\begin{figure}[h]
    \centering
    \vspace{-20pt}\includegraphics[width=0.98\linewidth]{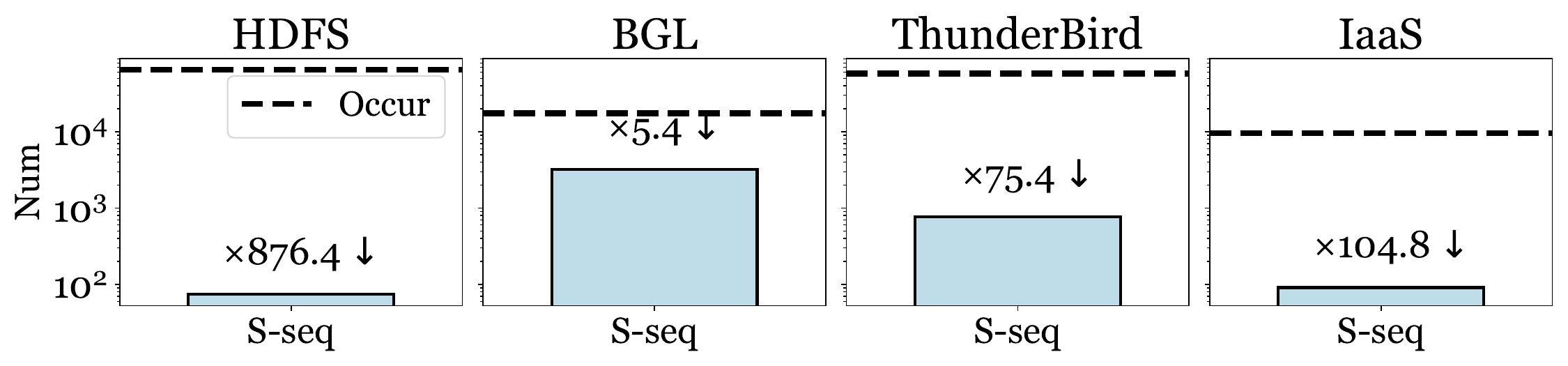}
    \vspace{-10pt}
    \caption{Re-usability of \statusseqs{}. Black dashed line: total occurrences of \statusseqs{}; bar: \# unique \statusseqs{}. $\downarrow$: size reduce w.r.t occurrences.}
    \vspace{-10pt}
    \label{fig:reusability}
\end{figure}

\color{black}

\subsection{Effectiveness of \app{} Hierarchy}
\label{sec:exp:hier_viz}


We perform a comprehensive analysis of the hierarchical structure introduced by \app{} across four datasets to evaluate its effectiveness and structural properties. 
Due to space constraints, we emphasize three key aspects:
(1)  benefits achieved by \app{}, including cardinality and resource reduction,  performance improvement;
(2) distribution of normal \appseqs{} within the training data across hierarchical levels; and
(3)  distribution of true anomalies detected by \app{} across hierarchical levels. 


\noindent\textbf{Benefits of \app{} Hierarchy. } We analyze the benefits of the \app{} hierarchy in terms of its data cardinality reduction, its  improvement of F-1 
for simple pattern matching detectors, its effectiveness compared to other hierarchies, and its facilitation of scalable LLM-based detection.  

$\bullet~$\textit{Cardinality Reduction.}
Figure~\ref{fig:cadinality} compares the number of training log sequences with the number of decomposed \statusseqs{} (the finest-grained \appseqs{}), and with patterns mined by PrefixSpan~\cite{prefixSpan}. We mine frequent, closed, and maximal patterns using a minimum support ratio of 0.5\% for large datasets—HDFS (28), BGL (231), Thunderbird (59)—and 10\% for the smaller IaaS dataset (36).
Despite these relatively high thresholds, PrefixSpan still suffers from severe pattern explosion~\cite{prefixSpan,pm_survey,zaki2001spade}, producing up to 8.7K$\times$ more patterns than the original data (IaaS). In contrast, \app{} retains every \statusseq{} that appears at least once yet still reduces the data space by 4.0$\times$–117.3$\times$. Moreover, unlike frequent patterns that must be re-mined from scratch whenever new data arrives, \appseqs{} form persistent, reusable knowledge units that enable incremental mining and  knowledge accumulation.  Figure~\ref{fig:reusability}  evaluates the reusability of \appseqs{} by comparing the number of unique \statusseqs{} against their total occurrences, revealing an average reduction of 5.4$\times$–876.4$\times$. \textit{The compactness of \statusseqs{} validates  reusability of atomic patterns as intrinsic behavioral units. }

\begin{table}[t]
    \caption{Detection results of  pattern matching with or without \app{} hierarchy. }
    \vspace{-5pt}
    \centering
    \resizebox{0.9\linewidth}{!}{
    \begin{tabular}{c|c|c|c|c}
    \toprule
         \textbf{Dataset} &\textbf{Method} & \textbf{Precision} & \textbf{Recall} & \textbf{F-1} \\\hline
         \multirow{2}{*}{HDFS} &Matching & 56.06 & 99.98 & 71.84\\\cline{2-5} 
         & \textbf{$\app_{SAE}$-P}& 76.65 & 99.99& \textbf{86.78}\\\hline
         \multirow{2}{*}{BGL}&Matching &35.80& 100.0& 52.72 \\\cline{2-5}
         & \textbf{$\app_{SAE}$-P}& 87.20& 99.99& \textbf{93.15}\\\hline
         \multirow{2}{*}{ThunderBird} &Matching & 12.48& 100.00& 22.20 \\\cline{2-5}
         & \textbf{$\app_{SAE}$-P}&45.33&100.00& \textbf{62.39} \\\hline
         \multirow{2}{*}{IaaS} &Matching &26.97 & 100.00 & 42.49 \\\cline{2-5}
         & \textbf{$\app_{SAE}$-P}&78.53& 100.00& \textbf{87.98}\\\bottomrule
    \end{tabular}}
\vspace{-15pt}
    \label{tab:comapre_PM}
\end{table}

$\bullet$~\textit{Improvement to Simple Pattern Matching Detector.}
To assess the practical benefit of the \app{} hierarchy, we adopt a simple pattern-matching baseline for the sequence level: a test sequence is flagged as abnormal if it does not appear in the training set, which may lead to many false positives. 
We compare it with \textbf{$\app_{SAE}$-P}, a variant of \app{} that employs pattern matching only as the \appseq{}-level detector (without LLM), where subscript \textit{SAE} denotes detection is enabled at the status, action, and entity levels. As shown in Table~\ref{tab:comapre_PM}, while the plain pattern-matching detector performs poorly, highlighting the task is non-trivial for simple detectors, however, $\app_{SAE}$-P achieves an absolute F-1 improvement of 14.94\%–45.49\%, demonstrating that hierarchical modeling substantially reduce problem complexity and enhances detection accuracy.

 \begin{table}[t]
 \vspace{-5pt}
      \caption{Detection performance using \app{} with different extracted semantic hierarchies.}
     \label{tab:hierarchy-comparison}
     \centering

     \resizebox{0.8\linewidth}{!}{
     \begin{tabular}{c|c|ccc}
     \toprule
      \multirow{2}{*}{Dataset}& \multirow{2}{*}{Hierarchy} & \multicolumn{3}{c}{Detection} \\\cline{3-5}
      && P(\%)$\uparrow$&R(\%)$\uparrow$&F-1(\%)$\uparrow$\\\hline\hline
     \multirow{3}{*}{HDFS}&hLDA &  58.42& 99.60& 73.64\\
     &Raptor & 54.18 & 72.89&62.16\\
     &\textbf{\app{}} &  \textbf{80.72} &\textbf{99.31} & \textbf{89.06} \\\hline
        \multirow{3}{*}{BGL}& hLDA&62.60&99.98&76.99\\
& Raptor&82.73&98.48&89.92 \\
& \textbf{\app{}} &\textbf{92.84} & \textbf{99.97}&\textbf{96.27}\\\hline
        \multirow{3}{*}{ThunderBird}& hLDA&23.95&100.00&38.65\\
        & Raptor&21.07&100.00&34.80\\
& \textbf{\app{}} &  \textbf{60.12}&100.00& \textbf{75.10} \\
        \bottomrule
     \end{tabular}}

\vspace{-10pt}
 \end{table}

$\bullet$~\textit{Effectiveness versus Other Hierarchies.} To benchmark the quality of the \app{} hierarchy in detection, we adapt  two popular hierarchical models of general text to log templates, and compare them to \app{}: (1) hLDA~\cite{hLDA}, a probabilistic topic model that infers hierarchical latent topics as word distributions, and (2) RAPTOR~\cite{RAPTOR}, an LLM-assisted framework that organizes LLM-generated text summaries into a multi-level hierarchy for retrieval. With all three extracted hierarchies, we  run \app{} detection using each of them respectively, on the three public datasets and report the results in Table~\ref{tab:hierarchy-comparison}.
Results show that the \app{} hierarchy consistently outperforms both baselines. While hLDA and RAPTOR induce topic-based hierarchies from word or embedding similarities with ambiguous data-dependent boundaries,  \app{}, in contrast, using an explicitly defined schema and constrained instantiation rules,
 yields more structurally grounded and reproducible patterns for anomaly detection.

\begin{figure*}[t]
    \centering

    \subfigure[\appseqs{} stored in training knowledge bases. 
    ]{
    \includegraphics[width=0.43\linewidth]{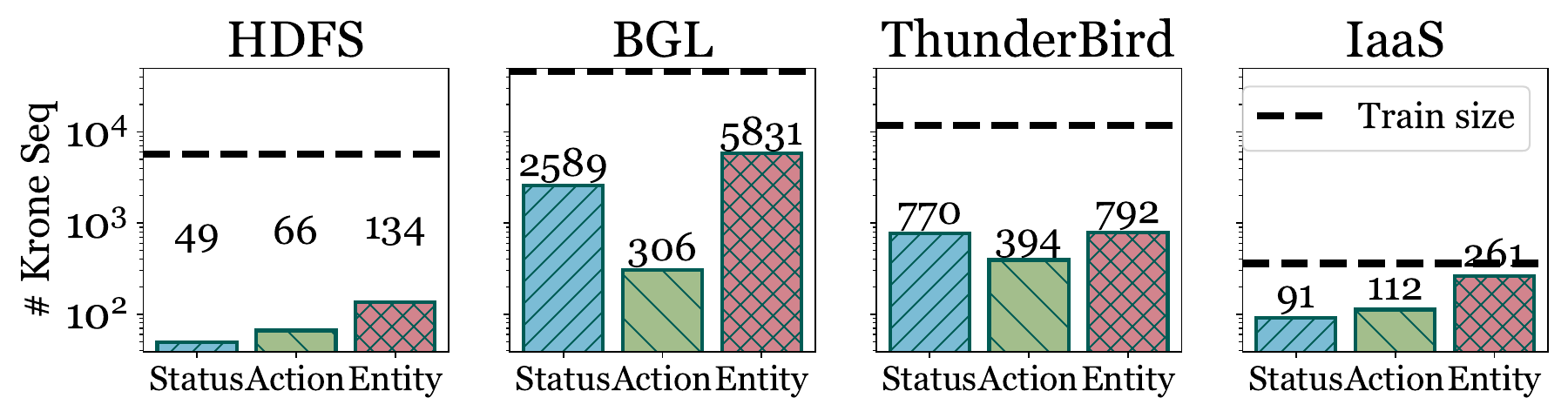}
\label{fig:hier_train}
}
    \subfigure[True anomalies detected by \app{} hierarchy.]{
    \includegraphics[width=0.43\linewidth]{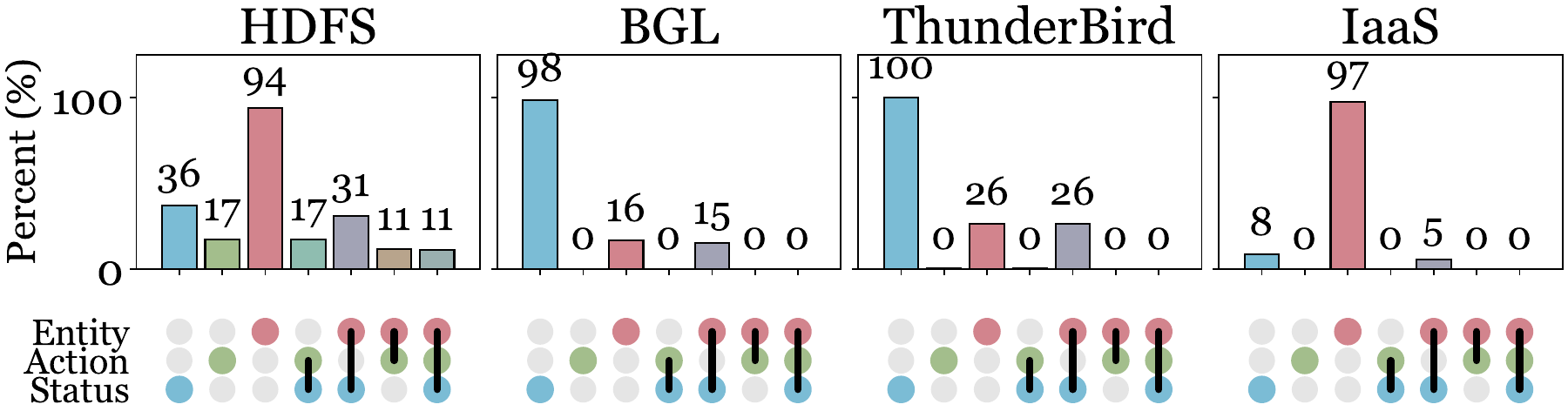}
\label{fig:hier_anomaly}
    }

\vspace{-5pt}
\caption{Visualization of \app{} Hierarchy. }
\label{fig:viz}
\vspace{-10pt}
\end{figure*}

\begin{figure}
    \centering
    \includegraphics[width=0.98\linewidth]{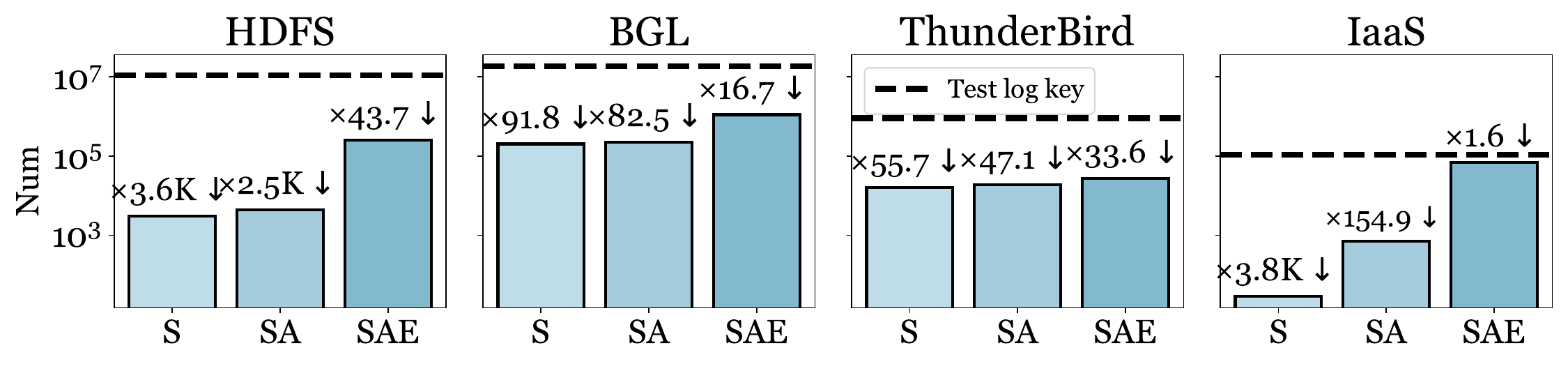}
    \vspace{-5pt}

    \caption{Resource reduction by \app{} (by \#log keys). S: log keys in \statusseqs{}; SA: log keys in \statusseqs{} and \actionseqs{}; SAE: log keys across all hierarchical \appseqs{}. }
    \label{fig:tokens}
    \vspace{-18pt}
\end{figure}

$\bullet$~\textit{Enabling Scalable LLM-based Detection.} Applying LLMs directly to massive log datasets is computationally and financially prohibitive. To estimate LLM-related costs, we treat each log key as  resource unit. Figure~\ref{fig:tokens} compares the total number of log keys in the original testing data with those in the decomposed test \appseqs{}. Whereas a naïve LLM-based approach  processes all log sequences and their constituent keys, hierarchical detection using \appseqs{} achieves substantial resource savings. Detection restricted to the Status level reduces the processing costs by 55.7$\times$–3.8K$\times$, while full hierarchical detection  achieves 1.6$\times$–43.7$\times$ reductions.


\color{black}

\noindent\textbf{Hierarchy of \appseqs{} in the Training Data.}
Figure~\ref{fig:hier_train}  shows the number of normal \appseqs{} decomposed from the training set {\it across all levels.} 
Decomposing log sequences into \appseqs{} reduces data cardinality and storage, \ie the number of total \appseqs{} versus the training set size, by 1 to 5-fold for less hierarchical datasets (Bytedance-IaaS, BGL) and 6 to 23-fold for more hierarchical datasets (ThunderBird, HDFS). This reduction highlights the \datamodel{} for efficient and effective anomaly detection. We  provide the decomposed data of the public-domain HDFS training set for use by others in the Appendix~\ref{appendix: exp}.

\color{blue}

\color{black}




\noindent\textbf{Case Study of Abstraction
Levels of Anomalies  Detected Across Different Data Sets.}
Unlike single-scale approaches, \app{} detects anomalies at three hierarchical levels—Entity, Action, and Status—capturing both fine-grained state transitions and high-level behavioral deviations.
To analyze how anomalies manifest across these levels, we disable the early-exit optimization 
(Section~\ref{sec:detection_framework}) so that \app{} continues evaluation even
after 
an anomaly on the low level  is detected.
Thus, each test sequence yields three 
predictions at all three levels. Figure~\ref{fig:hier_anomaly} shows the distribution of true anomalies detected at each level and their combinations.
In Thunderbird and BGL, nearly all anomalies (100\% and 98\%) occur at the Status level, suggesting short-range error patterns.
In contrast, most anomalies in HDFS (94\%) and Bytedance-IaaS (97\%) are at the Entity level, reflecting long-range behavioral deviations.
HDFS exhibits a more balanced distribution, indicating riche  multi-scale anomaly structures. Due to space limits,
{\it{examples of Status-, Action-, and Entity-level anomalies from HDFS}} are provided in the Figure~\ref{fig:example_anomalies}.

\subsection{Ablation Study of \app{} Innovations}
\label{sec:exp:Ablation}

\begin{figure}[t]
    \vspace{-10pt}
\subfigure{
\includegraphics[width=0.95\linewidth]{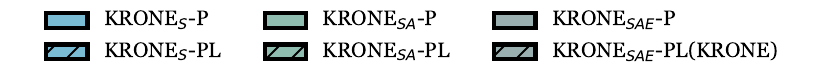}
}
\vspace{-10pt}

\subfigure{
\includegraphics[width=0.98\linewidth]{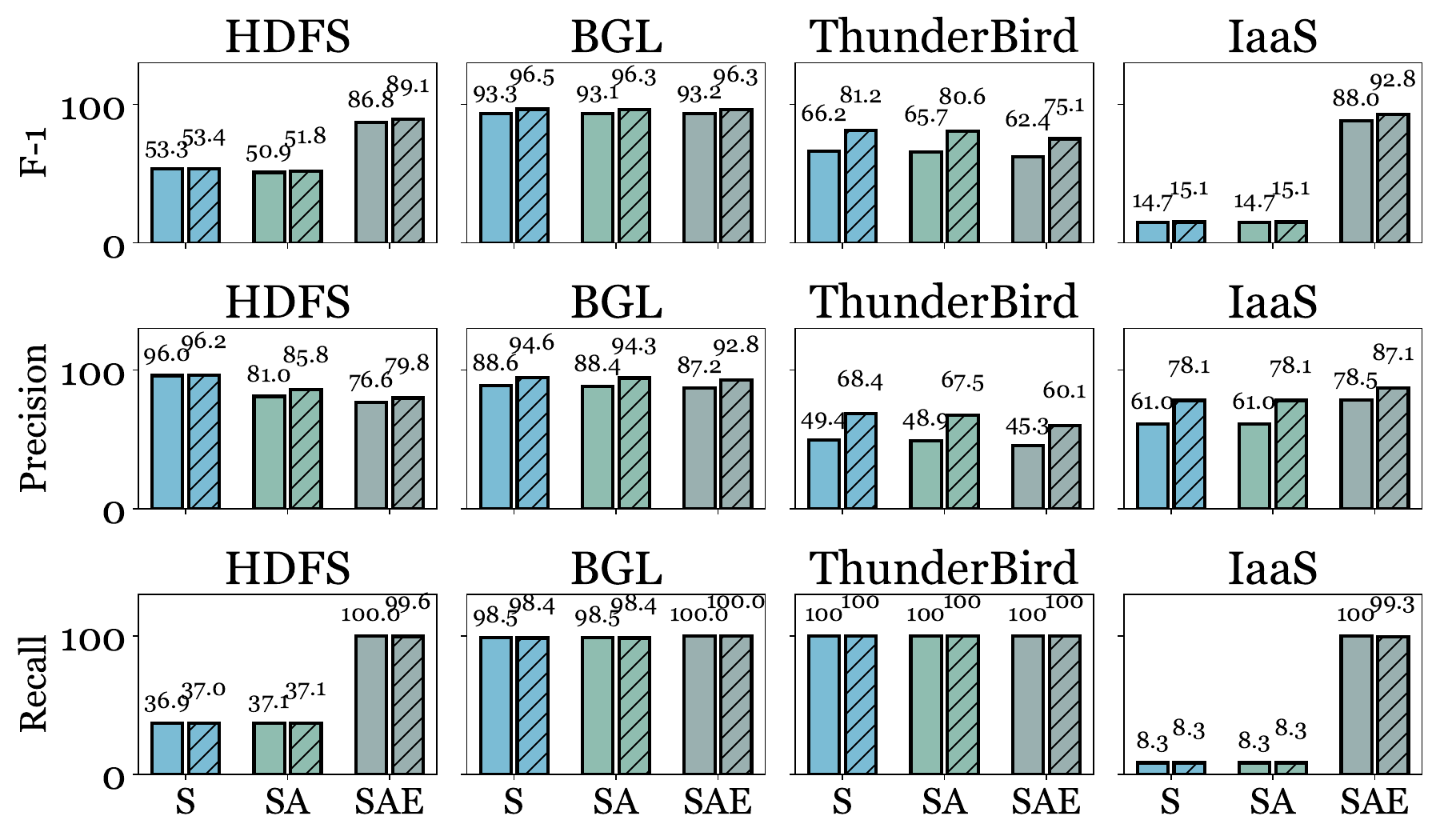}}
\vspace{-20pt}
\caption{Ablation Study of \app{}. $\app_{SAE}$-PL equals full \app{}. }
\label{fig:ablation}
\vspace{-18pt}
\end{figure}

We plug in  two  different detectors (Section \ref{sec:detection_framework}):
\textbf{\app{}-P}, which uses only pattern mining as the local context detector, and \textbf{\app{}-PL},  including both pattern mining and LLM detectors. For each, we create three level-specific variants: \textbf{$\app{}_S$}, \textbf{$\app{}_{SA}$}, and \textbf{$\app{}_{SAE}$}, by incrementally incorporating higher levels (Status, Action, Entity) into the detection process. Figure~\ref{fig:ablation} reports results for six \app{} variants and shows that the benefit of adding hierarchy levels depends on dataset characteristics.
For datasets with many high-level anomalies (HDFS, Bytedance-IaaS; Figure~\ref{fig:hier_anomaly}), incorporating more levels substantially improves F-1, from 53.4\% to 86.8\% on HDFS ($\app_S$-P vs. $\app_{SAE}$-P) and from 15.1\% to 92.8\% on Bytedance-IaaS ($\app_S$-PL vs. $\app_{SAE}$-PL).
For datasets dominated by low-level anomalies (BGL, ThunderBird), adding levels offers limited benefit; in ThunderBird, where 100\% of anomalies occur at the Status level, $\app_S$-PL achieves the best performance.
Across all level variants, -PL consistently outperforms -P, with F-1 gains from 0.2\% (HDFS, $\app_S$) to 15.1\% (ThunderBird, $\app_S$), validating the value of hierarchical structure and selective LLM integration.

\subsection{Effectiveness and Cost of LLM-Integration} 
\label{sec:exp:llm_effective}

 

\begin{figure}[t]
    \vspace{-2pt}
    \includegraphics[width=0.98\linewidth]{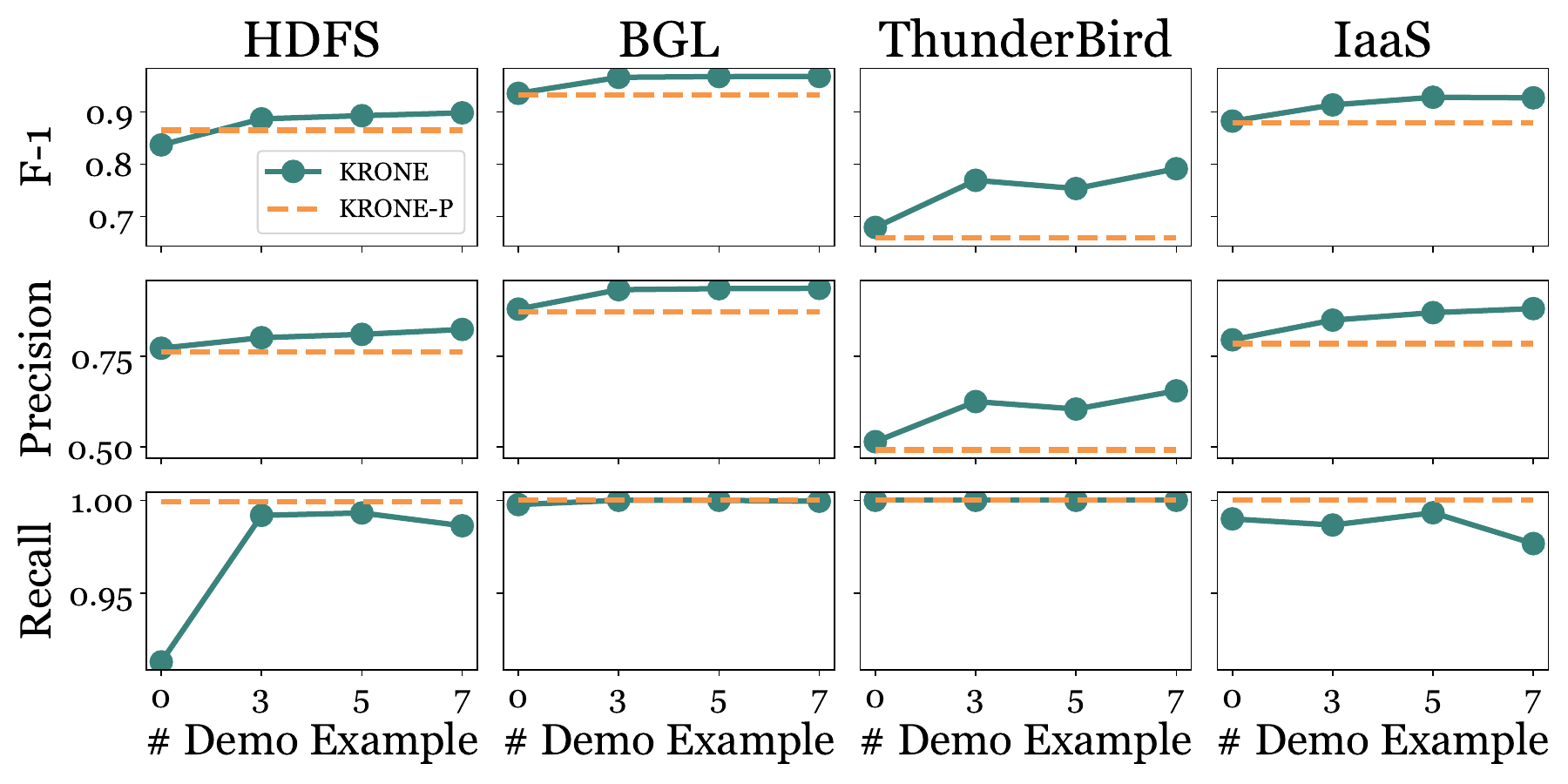}
    \vspace{-10pt}
    \caption{F-1 vs. varying demonstration example number.}
    \label{fig:demo_num}
    \vspace{-15pt}
\end{figure}

\noindent\textbf{Effectiveness - Varying Number of Demonstration Examples.} 
Figure~\ref{fig:demo_num} shows  F-1, Precision, and Recall of \app{} under numbers of (normal) examples $m$ ranging from 0 to 7. We also provide results of \app{}-P (layer subscripts SAE omitted for simplicity)  (dashed line). As expected, the precision and the consequent F-1 improve with more in-context examples across all datasets. When $m = 0$, zero-shot \app{} performs similarly to \app{}-P, which has no LLM integration, on most datasets, except for HDFS, where it under-detects anomalies and exhibits lower recall. These results show that pattern matching provides a high-recall baseline by reliably detecting symbolic abnormal \appseqs{}. 
When grounded with normal execution examples, LLM-based in-context reasoning improves precision by filtering false positives, while preserving recall. 
However, when anomalies lack obvious semantic cues (\eg HDFS), zero-shot LLM reasoning may under-detect anomalies and significantly reduce recall due to the absence of grounded normal-context understanding.

\noindent\textbf{Effectiveness - Varying LLM Integration Percentage in Testing.} To evaluate whether increased LLM integration improves detection quality, we measure the transition between \app{}-P, \ie $i=0\%$ with no LLM integration, and \app{}, $i=100\%$ with full integration. 
To this end, we split the test set into two subsets based on the ratio $i$ for the two phases. 

$\bullet~$\textit{Phase 1: LLM-Detector-Enabled} – In this phase, \app{} runs with the LLM detector activated only on the first $i\%$ portion of the unlabeled test data. LLM predictions (not labels) and summaries are stored in the test knowledge bases for potential reuse during later inference.
    
$\bullet~$\textit{Phase 2: Local-Test-Knowledge-Only} – In the second phase, LLM detector is disabled, instead  \app{}  relies only on the pattern matching detector and the test knowledge stored in the testing knowledge bases created during Phase 1.

\begin{figure}[t]

\includegraphics[width=0.9\linewidth]{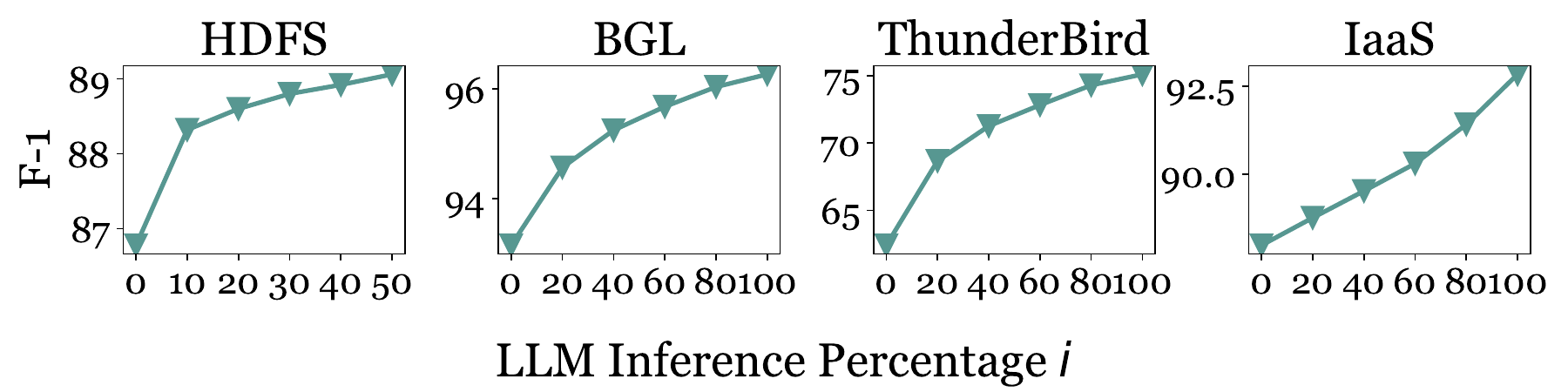} 

\vspace{-10pt}
\caption{F-1 vs. varying $i$ }
    \label{fig:llm_percent_f1}

\vspace{-15pt}

\end{figure}

For smaller datasets besides HDFS, we set $i\in[0,100]\%$, while for HDFS we cap $i$ at 50\% for cost control. 
We report results on the full test set. Figure~\ref{fig:llm_percent_f1} shows how detection performance varies with $i$, the percentage of test data processed by LLMs in Phase~1. Performance improves as more LLM involvement is allowed, but gains plateau—showing the effectiveness of caching results in the test knowledge base. Importantly, strong detection accuracy is achieved even when LLMs are applied to only a small fraction of logs, demonstrating that strategic, partial LLM use offers an effective balance between accuracy and resource cost.

\begin{figure}[t]

\includegraphics[width=0.99\linewidth]{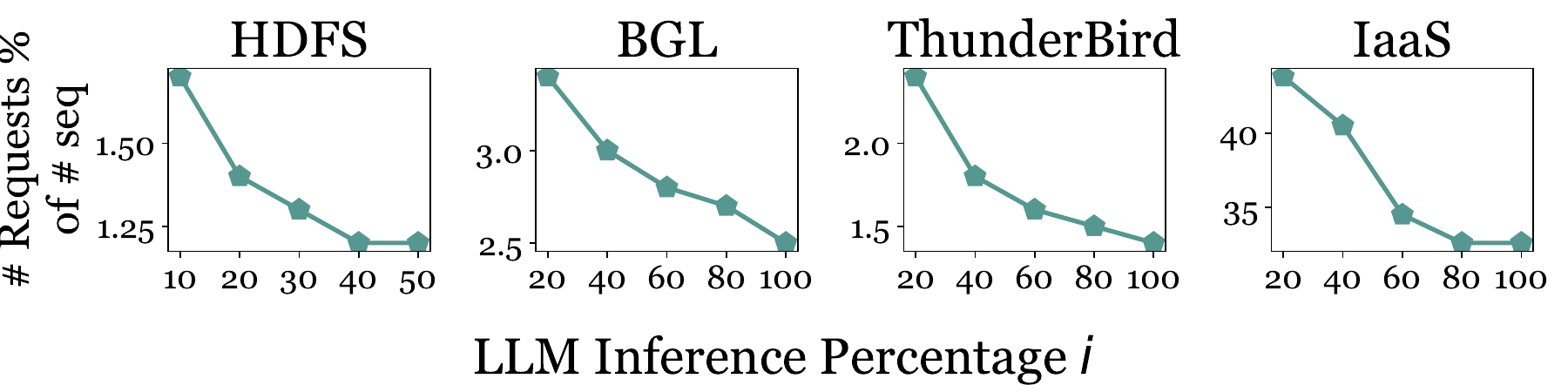}

\vspace{-10pt}
\caption{Request percentage (\%) of test size (phase 1) vs. varying $i$}
\label{fig:llm_percent_request_ratio}
\vspace{-15pt}
\end{figure}

\noindent\textbf{Costs for  LLM-Integration. } Under the same evaluation setup as Figure~\ref{fig:llm_percent_f1}, we quantify the actual LLM usage by measuring the ratio of LLM requests to the number of test sequences in Phase 1. As per  Figure~\ref{fig:llm_percent_request_ratio}, this ratio remains extremely low—only 1.1\%–3.3\% for the larger datasets (HDFS, BGL, Thunderbird). Aligning with the resource reductions observed in Figure~\ref{fig:tokens}, this observation demonstrates that \app{} enables  scalable LLM-based detection.

\section{Discussion}

\noindent\textbf{Future Works. }
As real-world systems evolve, inducing changes in log templates and execution behaviors, \app{} can support local hierarchy updates by appending new subtrees from newly observed templates and removing expired ones without recomputing the entire structure. 
\appseq{} occurrences provide a natural signal for detecting domain shifts of subtrees, which could guide automated hierarchy evolution. 
In addition, improving the robustness of \app{} to reduce dependence on log parsers and tolerate noisy parsed templates is another important direction for future research.

\vspace{0.02in}

\noindent\textbf{Limitations of \app{}.} 
\app{} relies on LLMs for the semantic hierarchy extraction. While our manual inspection and analysis did not reveal obvious hallucination, likely
due to \app{}’s well-defined schema that restricts the candidate space, we acknowledge that LLM hallucinations remain possible. This  could be  mitigated using advanced reasoning and reflection techniques~\cite{wei2023chainofthoughtpromptingelicitsreasoning,wang2023selfconsistencyimproveschainthought}. 
Broadly, deploying LLMs in industry-grade systems requires careful consideration of cost-effectiveness, operational efficiency~\cite{dai2024costeffectiveonlinemultillmselection, liu2024optllmoptimalassignmentqueries, li2025llmbanditcostefficientllm}, and concerns around safety and  data sovereignty~\cite{wang2024uniquesecurityprivacythreats, das2024securityprivacychallengeslarge, neel2024privacyissueslargelanguage, bondarenko2025sovereignlargelanguagemodels}. Another issue to consider is generality. The \datamodel{} applies to most logs that describe execution behaviors, but is less suitable when logs are hard to interpret (e.g., consisting only of codes, single words, or lacking behavioral semantics), in which case alternative strategies
 would be required.

\section{Conclusion}
We propose \app{}, a new paradigm for structure-aware log anomaly detection based on the \datamodel{}. 
By instantiating semantic hierarchies with LLMs, \app{} transforms flat log sequences into hierarchical execution chunks, enabling efficient multi-scale anomaly detection. 
Experiments on three public benchmarks and one industrial dataset show that \app{} consistently outperforms state-of-the-art methods. Beyond detection, \app{} offers insights into how anomalies manifest at different abstraction levels with explanations. 
We anticipate this work will inspire a new line of research on structured reasoning for log analytics, where traditional techniques and LLMs complement each other.

\section{AI-Generated Content Acknowledgment}
AI-based language tools (\eg ChatGPT) were used only for grammar, phrasing, and stylistic refinement. All conceptual content, analyses, experiments, and results were fully developed and validated by the authors, with no AI-generated scientific ideas, data, or findings.



\bibliographystyle{IEEEtran}

\bibliography{reference}

\clearpage
\appendix

\section{APPENDIX}

\subsection{Methodology}
\label{appendix: method}

\noindent\textbf{Summarization of \app{}.} Summarization is a standard technique for condensing long text while preserving essential information~\cite{summary_survey}, and is widely used in LLM-based tasks to handle long-context inputs. To enable in-context LLM detection for \appseqs{}, each \appseq{} requires a textual representation. However, directly concatenating log templates is redundant and quickly exceeds the context window, especially for higher-level \appseqs{} (e.g., an \entityseq{} spans the entire sequence). We therefore use summaries as compact \appseq{} representations, generated once and stored as metadata. To avoid unnecessary computation, we adopt a bottom-up strategy similar to divide-and-conquer summarization~\cite{divide_conquer_summary}: we first prompt the LLM to summarize each \statusseq{} using the concatenation of templates, then prompt the LLM to generate \actionseq{} summaries from their children \statusseqs{}'s summaries, and recursively do the same for \entityseqs{}. This keeps each LLM call short and focused, enables summary reuse across hierarchy levels, and substantially reduces LLM cost and recomputation.
\color{black}

\setlength{\textfloatsep}{0pt}
\begin{algorithm}[h]
\caption{\textsc{\partitionalgo{}}}
\label{algo:partition}
\begin{algorithmic}[1]
\Require A sequence of log keys $L=[k_i]$, the \app{} tree $T=(V,R)$, and level $l\in\{\text{entity},\text{action},\text{status}\}$.
\Ensure The list $\mathcal{X}$ of partitioned chunks of $L$ and the generated \appseq{} $S$ at level $l$.
\State $\mathcal{X} \gets [\ ],\quad S \gets [\ ]$ \Comment{Initialize outputs}
\State $x \gets [\ ]$ \Comment{Current temp chunk}
\ForAll{$k \in L$}
    \State $v \gets \Call{FindNode}{T, k, l}$ \Comment{Node for $k$ at level $l$}
    \If{$l>0 \ \wedge\  S \neq [\ ] \ \wedge\  v = S[-1]$}
        \State $\Call{ListAppend}{x, k}$ \Comment{Extend current chunk}
    \Else
        \If{$x \neq [\ ]$}
            \State $\Call{ListAppend}{\mathcal{X}, x}$ \Comment{Store completed chunk}
        \EndIf
        \State $x \gets [k]$ \Comment{Start new chunk}
        \State $\Call{ListAppend}{S, v}$ \Comment{Append node}
    \EndIf
\EndFor
\If{$x \neq [\ ]$}
    \State $\Call{ListAppend}{\mathcal{X}, x}$ \Comment{Flush last chunk}
\EndIf
\State \Return $\mathcal{X}, S$
\end{algorithmic}
\end{algorithm}

\setlength{\floatsep}{0.1cm}
\setlength{\textfloatsep}{0pt}

\smallskip
\noindent\textbf{Utility Algorithm \partitionalgo{}. } To support the recursive decomposition, we propose a utility algorithm \textsc{GenerateKroneSeq} (Algorithm~\ref{algo:partition}). Given a log sequence and a specified level, it returns a \appseq{} by mapping each log key to a node on the current level in the \app{} Tree. It also returns the corresponding list of log chunks $\mathcal{X}$. This utility algorithm 
is invoked at each level 
to enable segmentation. In the for-loop of Lines 3-9, given a log key $k$, the cost of finding the node and the list maintenance is both $O(1)$. Thus, let $|L|$ be the length of the input sequence, the complexity of Algorithm~\ref{algo:partition} is $O(|L|)$. 

\subsection{More Experimental Analysis}
\label{appendix: exp}

\begin{figure}[h]
    \subfigure[Nodes in \app{} Tree.]{
\includegraphics[width=0.92\linewidth]{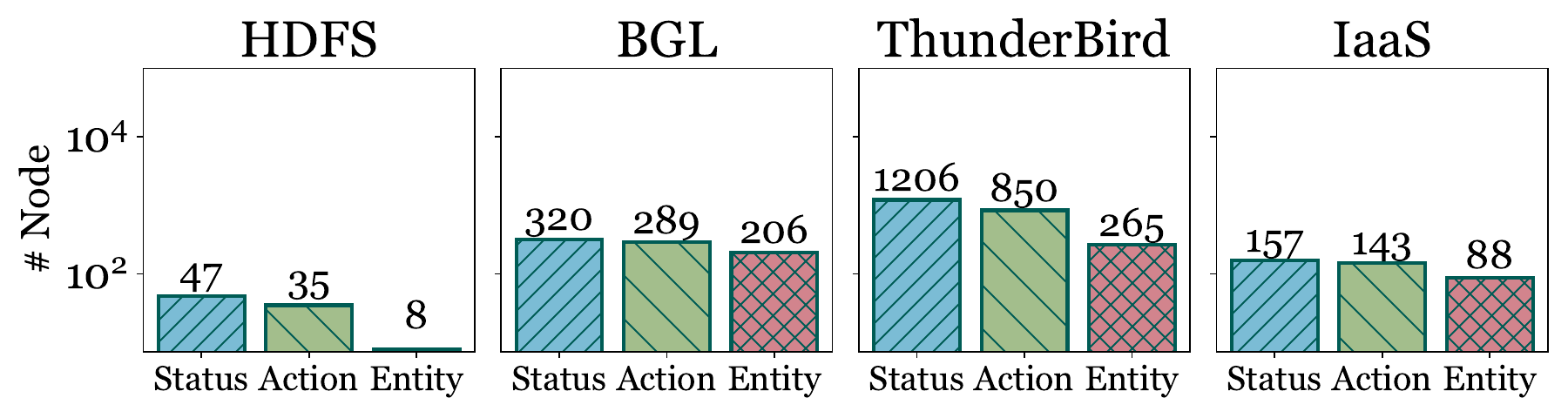}
\label{fig:hier_tree}
    }

\vspace{-10pt}
    \subfigure[\appseqs{} stored in testing knowledge bases. ]{
\includegraphics[width=0.92\linewidth]{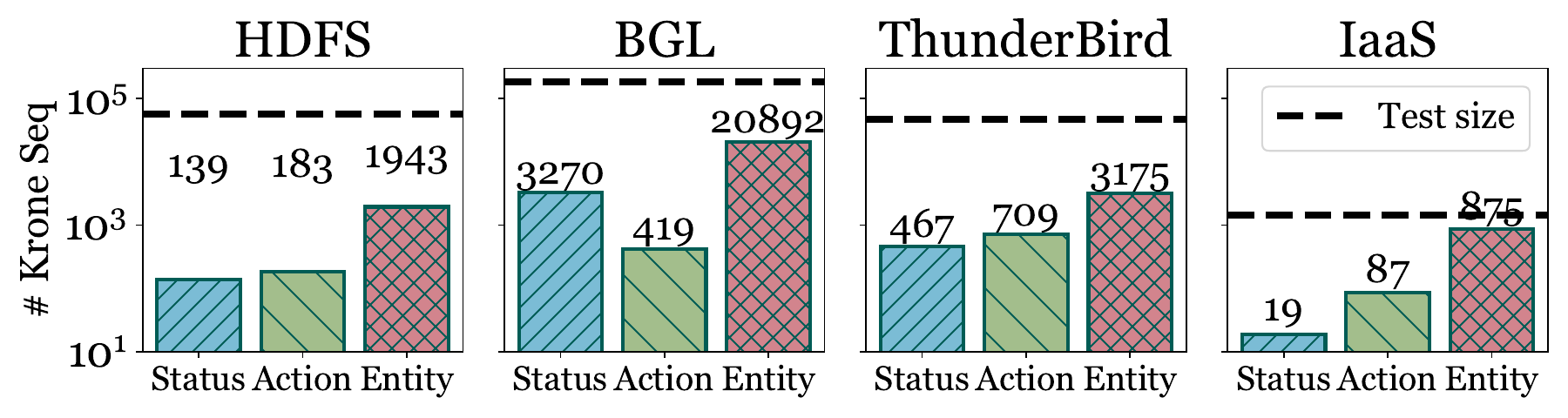}
\label{fig:hier_test}
}

\caption{More Visualization of \app{} Hierarchy. }
\label{fig:viz}
\end{figure}

\begin{figure}[h]
    \subfigure{
        \includegraphics[width=0.92\linewidth]{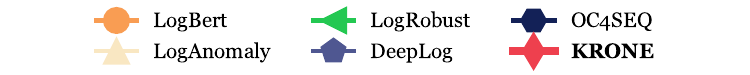}
    }%
    \vspace*{-4mm}
    \subfigure{
        \includegraphics[width=0.96\linewidth]{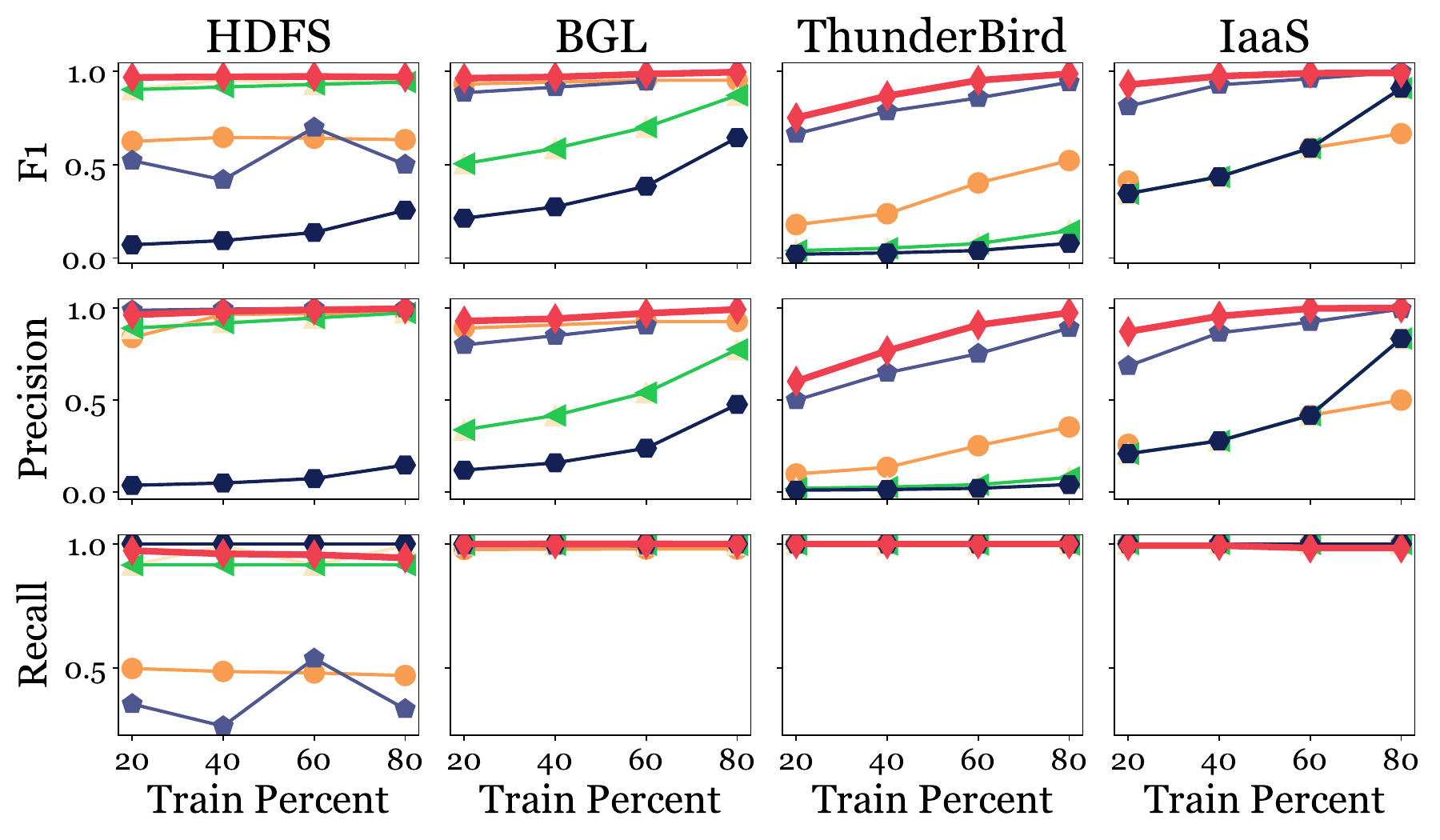}
    }%

    \caption{F-1 vs. training percentage.}
    \label{fig:train_percentage}

\end{figure}
\noindent\textbf{Hierarchy of \app{} Tree.} Figure~\ref{fig:hier_tree} shows the number of nodes on each level of the \app{} tree, \ie the number of Entities, Actions and Statuses extracted from the log templates. While the number of Status nodes always equals the number of log keys  in the training data, the number of Action and Entity nodes are usually fewer, showing the pyramid structure of the \app{} tree. Specifically, the BGL and Bytedance-IaaS datasets have more balanced levels,  while the HDFS and ThunderBird datasets have much fewer high-level nodes, indicating better hierarchical structure.

\smallskip
\noindent\textbf{Hierarchy of \appseqs{} in the Testing Data.} Figure~\ref{fig:hier_test} shows the testing set size versus the number of \appseqs{} decomposed from the testing set. Due to the larger test set compared to the training set, test knowledge bases have more overall \appseqs{} than the training knowledge bases. Similar with Figure~\ref{fig:hier_train}, by modular detection on each \appseq{}, \app{} achieves a significant reduction of data storage and computation overhead, \ie the number of total test \appseqs{} versus the test set size, ranging from 1 to 8-fold for the less hierarchical datasets (Bytedance-IaaS, BGL)  to 11 to 25-fold for more hierarchical datasets (ThunderBird, HDFS). While this reduction supports the local context pattern matching detector, its impact is even more significant for the LLM detector, particularly for optimizing LLM-related costs, as shown in Section~\ref{sec:exp:llm_effective}.

\begin{figure*}
    \centering
    \subfigure[Status Level Anomaly]{
\includegraphics[width=0.47\linewidth]{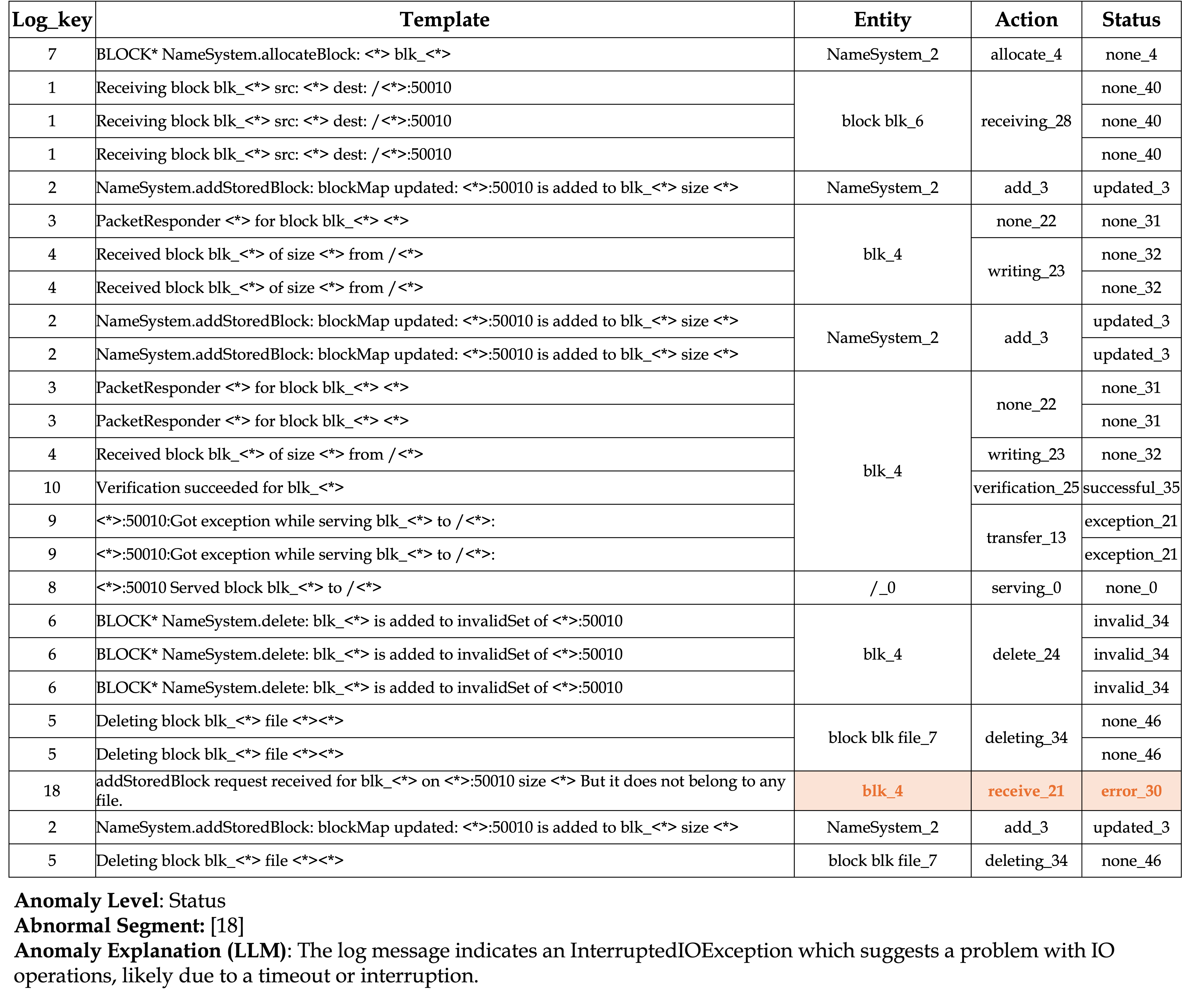}
\label{fig:status_anomaly}
    }
        \subfigure[Action Level Anomaly]{
\includegraphics[width=0.47\linewidth]{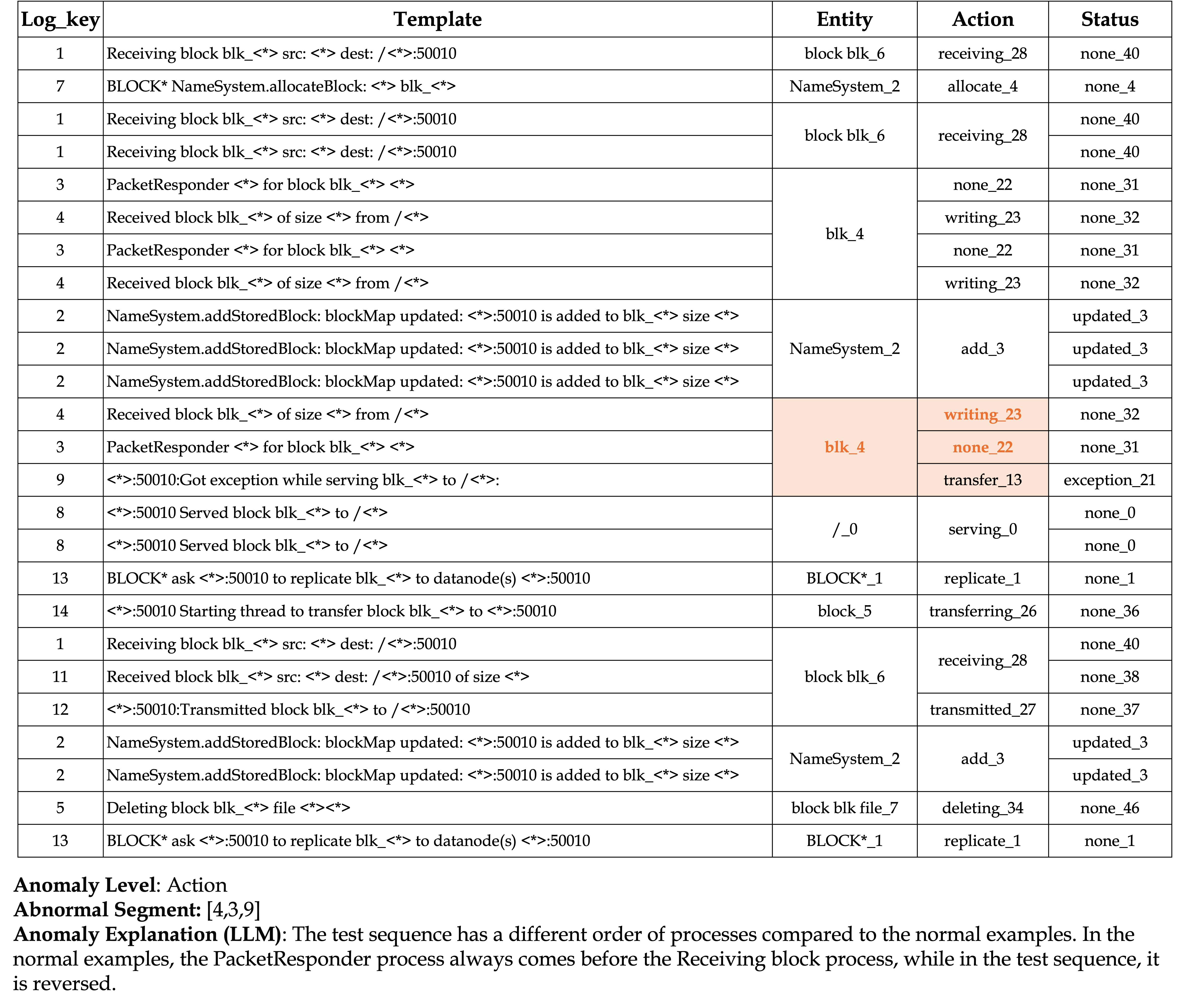}
\label{fig:action_anomaly}
    }
       \subfigure[Entity Level Anomaly]{
\includegraphics[width=0.6\linewidth]{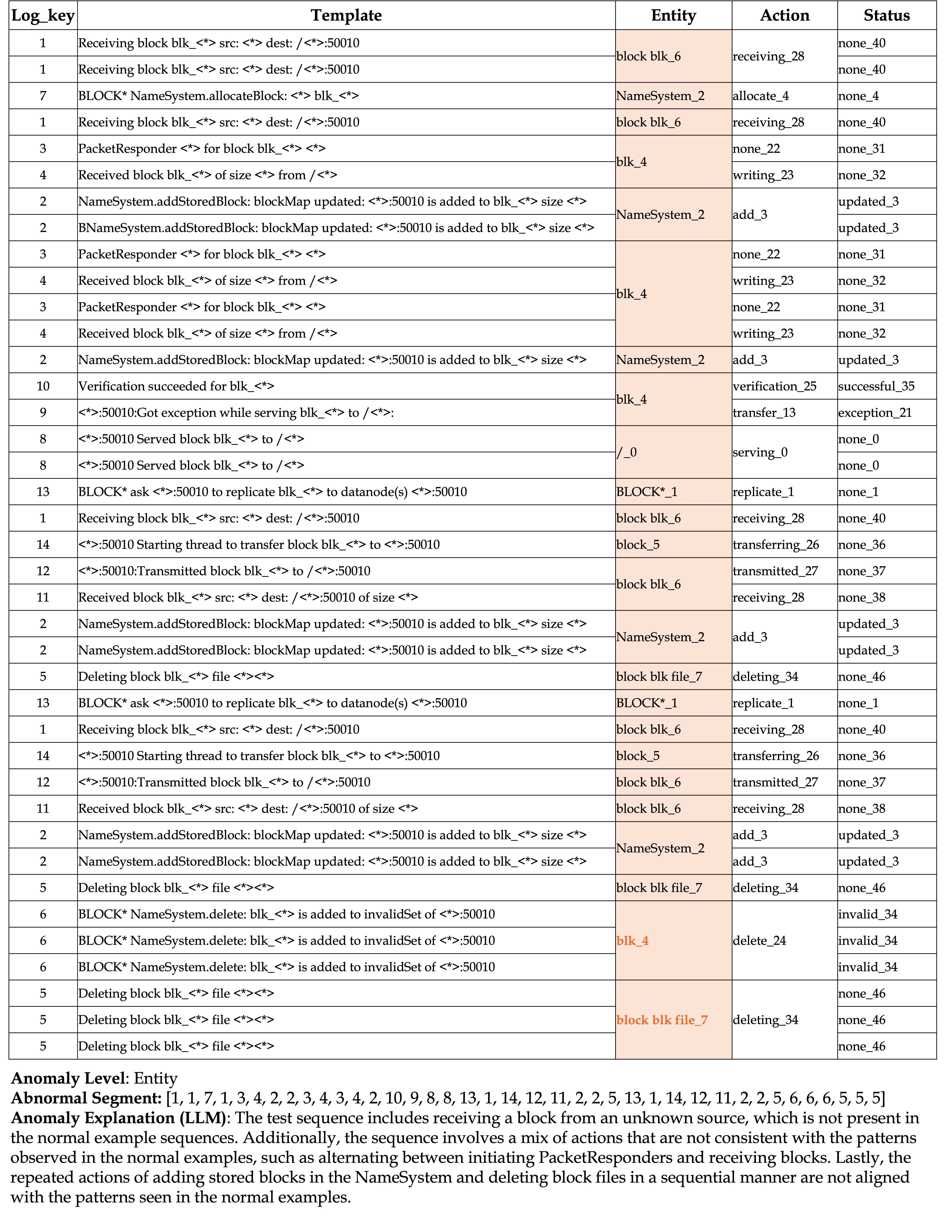}
\label{fig:entity_anomaly}
    }
    \vspace{-10pt}
        \caption{Examples of hierarchical anomalies detected by \app{} (HDFS dataset). }
    \label{fig:example_anomalies}
\end{figure*}

\smallskip
\noindent\textbf{Examples of Detected True Anomalies.} Figure~\ref{fig:example_anomalies} presents three representative true anomalies identified by \app{} at different hierarchical levels—\textit{status}, \textit{action}, and \textit{entity}.
At each level, \app{} not only localizes the abnormal segment but also generates an interpretable explanation through the integrated LLM.
For the status-level case, \app{} detects that status~$[error\_30]$ is abnormal for action~$receive\_21$.
For the action-level case, it identifies the action sequence~$[writing\_23, none\_22, transfer\_13]$ as inconsistent with the expected workflow of entity~$blk\_4$, with the LLM noting the reversed order of the packet-response and receiving processes.
At the entity level, \app{} highlights an anomalous entity sequence that spans the entire log, revealing a long-range and compound deviation across components.

\smallskip
\noindent\textbf{Varying the Training Percentage.} To explore prevalent trends of using deep learning for log anomaly detection \cite{landauer2023deep, le2022log}, we compare \app{} to  DL methods under varying train-test split ratios, ranging from 2:8, 4:6, 6:4 to 8:2. 

As in Figure~\ref{fig:train_percentage}, with more training data, as expected, we observe a common trend of performance increase among all methods across datasets.  Although some methods perform comparably to \app{}, \eg \logrobust{} and \loganomaly{} on HDFS, \deeplog{} on BGL, ThunderBird, and Bytedance-IaaS,  there is no consistent runner-up across datasets. Such observation suggests that log datasets may exhibit different characteristics in terms of pattern lengths, anomaly types, \etc, which may benefit from appropriate detection methods. Impressively, \app{} outperforms the comparative methods under varying training percentages across all datasets due to its adaptive granularity and semantic augmentation.

\begin{table}[t]
    \centering
        \caption{\app{}'s average detection time per sequence, with different detection settings.}
    \resizebox{1.0\linewidth}{!}{
    \begin{tabular}{c|c|c|c}
    \toprule
        Dataset & Detection components & average over & avg time (sec)\\\hline
        \multirow{4}{*}{HDFS} & PM only & whole& 1.5e-04\\
        & PM+knowledge & whole& 1.5e-04\\
        & PM+knowledge+LLM& applied& 6.3e-01\\
        & PM+knowledge+LLM &whole& 1.1e-02\\\hline
            \multirow{4}{*}{BGL} & PM only & whole& 3.9e-04\\
    & PM+knowledge & whole& 3.9e-04\\
    & PM+knowledge+LLM& applied& 1.6e+00\\
    & PM+knowledge+LLM (overall)&whole&5.3e-02 \\\hline
            \multirow{4}{*}{ThunderBird} & PM only & whole& 1.8e-04\\
    & PM+knowledge & whole& 1.8e-04\\
    & PM+knowledge+LLM  & applied&9.4e-01 \\
    & PM+knowledge+LLM   &whole& 2.4e-02\\\hline
        
    \end{tabular}}

    \label{tab:latency}
\end{table}

\smallskip
\noindent\textbf{Latency of \app{}. } We evaluate the average detection time per sequence under four settings on HDFS, BGL, and ThunderBird (Table~\ref{tab:latency}): (i) pattern matching only, (ii) pattern matching with knowledge lookup, (iii) pattern matching + knowledge + LLM (applied only to sequences triggering LLM), and (iv) the full pipeline averaged over the entire test set.
Results show that pattern matching and knowledge lookup incur negligible overhead ($\sim10^{-4}$s per sequence), indicating that \app{} can be configured for real-time or near–real-time screening. For the small subset of sequences that invoke LLM reasoning, latency is dominated by the external API (0.63–1.6s per sequence), which is acceptable for offline or human-in-the-loop diagnosis. Since LLMs are triggered for only a tiny fraction of sequences, the end-to-end average latency over the full workload remains low (less than 0.1s per sequence), preserving high throughput while enabling deeper semantic analysis when needed.

\subsection{More Discussion}
\label{appendix: discussion}

\noindent\textbf{\app{} Compared to Trace-based and Graph-based Diagnosis. } 
Trace- and graph-based diagnosis methods assume access to explicit execution structure, such as request traces, spans, or predefined service dependency graphs~\cite{pinpoint,magpie,sigelman2010dapper,zhang2019graph,zhou2023groot}. These approaches localize faults by reasoning over known component relationships and propagation paths. In contrast, log anomaly detection assumes only raw, unstructured log streams, without trace IDs or dependency graphs. \app{} follows this latter setting and requires only logs as input.
While traces and dependency graphs explicitly encode execution and propagation structure, such instrumentation is often unavailable or incomplete in practice. Logs are ubiquitous but lack structure, limiting traditional log-based diagnosis. \app{} bridges this gap by reconstructing a semantic execution hierarchy directly from logs, enabling structured reasoning without relying on tracing infrastructure.

\smallskip
\noindent\textbf{Impact of Log Parsing Errors. }In practice, parsers mainly produce three types of errors:
(i) parameter misclassification (masking meaningful tokens),
(ii) template fragmentation (splitting one event into multiple templates), and
(iii) template collision (merging distinct events into one template).
The first two are largely benign for \app{}: missing entity/action/status tokens can be handled via fallback candidates or a unique “None” label without breaking the hierarchy, while fragmentation mainly reduces \appseq{} reuse and thus affects efficiency rather than correctness~\cite{10.1007/s10664-022-10214-6}. The only truly fatal error—both for \app{} and for log-based anomaly detection in general—is template collision, which irreversibly erases anomaly signals by merging normal and anomalous events into the same template. Prior studies show that parsing errors can degrade anomaly detection~\cite{7579781}, but comprehensive evaluations~\cite{zhu2019toolsbenchmarksautomatedlog,7579781,10.1007/s10664-022-10214-6} also report that modern parsers—especially Drain~\cite{drain}—achieve very high accuracy on the public datasets used here (HDFS: 0.998, BGL: 0.963, ThunderBird: 0.955). This supports that parsing noise is limited in these settings and justifies the widespread use of template-based preprocessing by most SOTA methods~\cite{deeplog,logbert,meng2019loganomaly,logrobust,pluto,RT-log}, which we also follow in \app{}.

\end{document}